\documentclass[twocolumn, trackchanges]{aastex631}

\newcommand{\rproc}{$r$-process}

\newcommand{\emhstar}{2MASS J09544277+5246414}
\newcommand{\emhstarShort}{J0954+5246}
\newcommand{\annastar}{HE 1523-0901}
\newcommand{\hillstar}{CS 31082-001}
\newcommand{\cs}{CS 29497-004}
\newcommand{\vinnistar}{RAVE J203843.2–002333}
\newcommand{\vinnistarShort}{J2038-0023}
\newcommand{\eTeff}{$T_\mathrm{eff}$}
\newcommand{\logg}{$\log\ g$}
\newcommand{\vmicro}{$\xi$}
\newcommand{\loggf}{$\log\ gf$}
\newcommand{\utwo}{\ion{U}{2}}
\newcommand{\thtwo}{\ion{Th}{2}}
\newcommand{\epsU}{$\log\epsilon$(U)$\ =\ $}
\newcommand{\eps}{$\log\epsilon$}
 
\usepackage{comment}
\usepackage{footnote}
\usepackage{graphicx}
\usepackage{amsmath}
\usepackage{amssymb}
\usepackage{hyperref,enumerate}
\usepackage{natbib}
\usepackage[T1]{fontenc}

\shorttitle{U Abundances with New \utwo\ Lines}
\shortauthors{Shah et al.}
\graphicspath{{./}{figures_main/}}

\begin{document}

\title{Uranium Abundances and Ages of $R$-process Enhanced Stars with Novel U II Lines\footnote{ Some of the data presented herein were obtained at the W. M. Keck Observatory, which is operated as a scientific partnership among the California Institute of Technology, the University of California and the National Aeronautics and Space Administration. The Observatory was made possible by the generous financial support of the W. M. Keck Foundation. Additionally, this work is based on observations made with ESO Telescopes at the La Silla Paranal Observatory under programme IDs 275.D-5028(A), 077.D-0453(A), and 165.N-0276(A). This paper also includes data gathered with the 6.5 meter Magellan Telescopes located at Las Campanas Observatory, Chile.}}

\author[0000-0002-3367-2394]{Shivani P. Shah}
\affiliation{Department of Astronomy, University of Florida, 211 Bryant Space Science Center, Gainesville, FL 32601, USA}

\author[0000-0002-8504-8470]{Rana Ezzeddine}
\affiliation{Department of Astronomy, University of Florida, 211 Bryant Space Science Center, Gainesville, FL 32601, USA}

\author[0000-0002-4863-8842]{Alexander P. Ji}
\affiliation{Department of Astronomy \& Astrophysics, University of Chicago, 5640 S Ellis Avenue, Chicago IL 60637, USA}
\affiliation{Kavli Institute for Cosmological Physics, University of Chicago, Chicago IL 60637, USA}

\author[0000-0001-6154-8983]{Terese Hansen}
\affiliation{Department of Astronomy, Stockholm University, AlbaNova University Centre, SE-106 91 Stockholm, Sweden}

\author[0000-0001-5107-8930]{Ian U.\ Roederer}
\affiliation{Department of Astronomy, University of Michigan, 1085 S.\ University Ave., Ann Arbor, MI 48109, USA}
\affiliation{Joint Institute for Nuclear Astrophysics -- Center for the Evolution of the Elements (JINA-CEE), USA}

\author[0000-0001-6003-8877]{M\'arcio Catelan}
\affiliation{Instituto de Astrofísica, Facultad de Física, Pontificia Universidad Católica de Chile, Av. Vicu\~na Mackenna 4860, 7820436 Macul, Santiago, Chile}

\affiliation{Millennium Institute of Astrophysics, Nuncio Monseñor Sotero Sanz 100, Of. 104, 7500000 Providencia, Santiago, Chile}

\affiliation{Centro de Astroingeniería, Facultad de Física, Pontificia Universidad Católica de Chile, Av. Vicu\~na Mackenna 4860, 7820436 Macul, Santiago, Chile}

\author[0000-0002-3855-3060]{Zoe Hackshaw}   
\affiliation{Department of Astronomy, University of Texas at Austin, 2515 Speedway, Austin, Texas 78712-1205, USA}

\author[0000-0002-5463-6800]{Erika M. Holmbeck}
\affiliation{The Observatories of the Carnegie Institution for Science, 813 Santa Barbara St, Pasadena, CA 91101, USA}
\affiliation{Hubble Fellow}
%\affiliation{Joint Institute for Nuclear Astrophysics - Center for Evolution of the Elements, USA}

\author[0000-0003-4573-6233]{Timothy C. Beers}
\affiliation{Department of Physics and Astronomy, University of Notre Dame, Notre Dame, IN 46556, USA}
\affiliation{Joint Institute for Nuclear Astrophysics -- Center for the Evolution of the Elements  (JINA-CEE), USA}

\author[0000-0002-4729-8823]{Rebecca Surman}
\affiliation{Department of Physics and Astronomy, University of Notre Dame, Notre Dame, IN 46556, USA}
\affiliation{Joint Institute for Nuclear Astrophysics -- Center for the Evolution of the Elements  (JINA-CEE), USA}

\begin{abstract}
The ages of the oldest stars shed light on the birth, chemical enrichment, and chemical evolution of the Universe. Nucleocosmochronometry provides an avenue to determining the ages of these stars independent from stellar evolution models. The uranium abundance, which can be determined for metal-poor \rproc\ enhanced (RPE) stars, has been known to constitute one of the most robust chronometers known. So far, U abundance determination has used a \textit{single} \ion{U}{2} line at $\lambda3859$\,\AA. Consequently, U abundance has been reliably determined for only five RPE stars. 
%Here, we present high resolution spectroscopy of two novel U II lines at $\lambda4050$ and $\lambda4090$\,\AA. 
Here, we present the first homogeneous U abundance analysis of four RPE stars using two novel \utwo\ lines at $\lambda4050$\,\AA\ and $\lambda4090$\,\AA, in addition to the canonical $\lambda3859$\,\AA\ line. We find that the \utwo\ lines at $\lambda4050$\,\AA\ and $\lambda4090$\,\AA\ are reliable and render U abundances in agreement with the $\lambda3859$ U abundance, for all the stars. We, thus, determine revised U abundances for RPE stars, \emhstar, \vinnistar, \annastar, and \hillstar, using multiple \utwo\ lines. We also provide nucleocosmochronometric ages of these stars based on the newly derived U, Th, and Eu abundances. The results of this study open up a new avenue to reliably and homogeneously determine U abundance for a significantly larger number of RPE stars. This will, in turn, enable robust constraints on the nucleocosmochronometric ages of RPE stars, which can be applied to understand the chemical enrichment and evolution in the early Universe, especially of \rproc\ elements.

\end{abstract}


\section{Introduction}

Ages of the oldest stars aid our understanding of chemical enrichment and evolution in the early universe, the assembly history of our Galaxy \citep[e.g.,][]{Marin-Franch+2009, Bonaca+H32020, Xiang&Rix2022, Rix+2022, Buder2022}, and the age of the Universe \citep[e.g.,][]{Bond2013_HD140283, VandenBerg2014_AgesIsochrone, Jimenez2019_H0StellarAges, Valcin2020_GCAges, Abdalla2022_cosmology}. %However, stellar ages can not be measured, but only inferred \citep[see][for a review]{Soderblom2010_RevStellarAges}. 
Most techniques used to infer stellar ages, including isochrone-placement and asteroseismology, depend on a detailed understanding of low-metallicity stellar-evolution models, which is challenging and still evolving \citep[e.g.,][]{Migilo20113_scalingRelationsRedGiants, Epstein2014, Joyce2015_GCAges, Tayar2017_MixingLength, catelan2018_IAU, Valentini2019_AgesMetalPoorStars}. On the other hand, nucleocosmochronometry, a radioactive-dating technique, offers an independent avenue to determining the ages of some of the oldest stars \citep{Soderblom2010_RevStellarAges, catelan2018_IAU}.
\par
Nucleocosmochronometry uses the decay of long-lived actinides, uranium ($^{238}$U; $\tau_{1/2} = 4.47$ Gyr) and thorium ($^{232}$Th; $\tau_{1/2} = 14.05$ Gyr), to estimate the ages of metal-poor stars \citep{Francois+1993, Cowan1991Rev,  Cayrel2001_CS31082, FrebelKratz2009_Nucleocosmochronometry}. U and Th are created solely via the rapid-neutron capture process (\rproc) \citep{B2FH, Cameron1957}. Therefore, \rproc-enhanced (RPE) stars, classified as having [Eu/Fe] $> +0.3$\footnote{$\mathrm{[A/B]} = \log(N_A/N_B)_\mathrm{Star} - \log({N_A/N_B})_\mathrm{Solar}$, where $N$ is the number density of the element.} \citep{Beers&Christlieb2005Rev, rpa4}, have been some of the best candidates for employing nucleocosmochronometry \citep{Frebel18_rev}. RPE stars are typically metal-poor ([Fe/H] $\lesssim -1.5$; \cite{Frebel18_rev}), offering the ability to detect the weak absorption lines of U and Th in their spectra. Moreover, the \rproc\ enrichment of RPE stars is the result of only a few \rproc\ nucleosynthetic events, dismissing the need for galactic chemical enrichment models in nucleocosmochronometry \citep{Frebel18_rev, ArnouldGoriely2020_Rev}. In practice, the absolute ages of the stars are determined by using the observed present-day abundance ratios and theoretical production ratios (PRs) of U and/or Th to coproduced \rproc\ elements e.g., U/Th, U/X, and Th/X, where X refers to a lighter stable \rproc\ element, such as Eu, Os, and Ir \citep[e.g.,][]{Cowan1997_1stcosmochronometry, Cowan1999_2ndcosmochronometry_refined, Cayrel2001_CS31082, HillCS31082_2002, FrebelHE1523_2007, PlaccoJ2038_2017}.
\par
One of the major systematic uncertainties in nucleocosmochronometry is the PRs of Th and U to lighter \rproc\ elements like Eu \citep{Goriely2001_multieventrprocmodel, Schatz2002}. This issue has been prominently highlighted by the negative Th/Eu stellar ages obtained for 30\% of RPE stars, termed as actinide-boost stars \citep{Roederer+2009, Mashonkina2014, holmbeckJ0954_2018}. The negative stellar ages are a result of the observed Th/Eu abundance ratios being higher than the Th/Eu PRs predicted by current \rproc\ models. More generally, the application of the Th/Eu chronometer to RPE stars has led to a large range in ages from 21 Gyr to -9 Gyr, even though these stars are metal-poor \citep{Ji2018_actinidedeficient, holmbeckJ0954_2018}. These anomalies have indicated that the astrophysical conditions of \rproc\ nucleosynthetic events may be varying event-to-event, with the PRs of actinides to lighter \rproc\ elements sensitive to these changes. Consequently, no one set of Th/X and U/X PR may be applicable to all RPE stars \citep{Holmbeck2019b_samesite}.
\par
On the other hand, the U/Th chronometer results in high-fidelity stellar-age estimates even for the actinide-boost stars \citep{Cayrel2001_CS31082, HillCS31082_2002, FrebelHE1523_2007, PlaccoJ2038_2017, holmbeckJ0954_2018}. Since U and Th have similar nuclear masses and are synthesized along similar reaction channels during the \rproc, their PR is robust to major shortcomings of \rproc\ models \citep{Arnould&takahashi1999, Goriely&Clerbaus1999, Schatz2002}. Additionally, any variations in the \rproc\ astrophysical conditions are expected to impact the actinides, U and Th, equally, so that their PR is generally constant across all \rproc\ nucleosynethetic events \citep[e.g.,][]{Holmbeck2019b_samesite}, with the uncertainty in the predicted value of the PR largely due to the unknown nuclear data of the neutron-rich actinides \citep{Holmbeck2019a_ADM, Lund+2022}.
\par
However, it is particularly challenging to reliably determine U abundance. So far, in the context of nucleocosmochronometry, U abundance has been confidently determined for only five highly \rproc\ enhanced ([Eu/Fe] > +0.7) stars: namely \hillstar\ \citep{Cayrel2001_CS31082, HillCS31082_2002}, \annastar\ \citep{FrebelHE1523_2007}, \emhstar\ \citep{holmbeckJ0954_2018}, \vinnistar\ \citep{PlaccoJ2038_2017}, and \cs\ \citep{Hill2017_cs29497}. Canonically, and also in the case of these five stars, a $single$ \utwo\ line at $\lambda3859$ {\AA} has been used to determine the U abundance of RPE stars. This line is blended with the wing of a strong \ion{Fe}{1} line and a poorly-constrained CN feature, rendering a reliable U abundance determination challenging. Moreover, the proximity to the CN feature limits the study of U in stars with strong C enhancements. Interestingly, the U abundance of the Przybylski star (HD 101065), a chemically peculiar star, has been determined using 17 \utwo\ transitions \citep{Shulyak2010_PrzybylskiStarUpdate_UIILines}.
\par
In this study, we have homogeneously determined the U abundance of four highly RPE stars using two novel \utwo\ lines at $\lambda4050$ and $\lambda4090$\,\AA, in addition to the canonical $\lambda3859$\,\AA\ line. We revisit the U abundances of  \emhstar\ (hereafter \emhstarShort), \vinnistar\ (hereafter \vinnistarShort), \annastar, and \hillstar\ to investigate the utility of the two new \utwo\ lines. This work is aimed at serving as a benchmark test for the reliable determination of U abundances and subsequently stellar ages of RPE stars using multiple \utwo\ lines, in an effort to advance the field of nucleocosmochronometry. 
\par
Hereafter, this paper is organized as follows: Section~\ref{obs_reduction} describes the observations and data reduction of the stars. Section~\ref{sec:stellarparams} discusses their atmospheric stellar-parameter estimates. The chemical-abundance analysis of the pertinent elements, including U, Th, and Eu, is described in section~\ref{sec:chemAbund}. In Section~\ref{sec:age_estimate}, we present the nucleocosmochronometric ages of the stars using the newly derived U, Th, and Eu abundances. Finally, in section~\ref{sec:discussion}, we discuss our results and in section~\ref{sec:conclusions}, we present the main conclusions of this work.

\begin{deluxetable*}{ccccccc}
\tablenum{1}
\tablecaption{Spectral Data Properties\label{tab:observationLog}}
\tablehead{
\colhead{Star Name} & \colhead{Telescope/}  & \colhead{Wavelength} & \colhead{Slit}  & \colhead{Resolving Power} & \colhead{Total} & \colhead{S/N pix$^{-1}$} \\
\colhead{} & \colhead{Instrument} & \colhead{Range (\AA)} & \colhead{Width}  & \colhead{($\Delta\lambda/\lambda$)} & 
\colhead{Exposure (h)} & \colhead{at 4050\,\AA}
}
%\decimalcolnumbers
\startdata
\emhstar & Keck/HIRESb  & 3600-6800 & $0.40''$ &  $86,\!600$ & $6.6$  & $200$\\
\vinnistar & Magellan/MIKE & 3200-9900 & $0.35''$ & $83,\!000$ &  $15.6$  &  $175$\\
\annastar & VLT/UVES  & 3758-4990  &$0.45''$  & $75,\!000$ & $3.0$  & $200$ \\
\hillstar & VLT/UVES &  3800-5100 & $0.45''$ & $70,\!000$ & $2.0$  & $125$
%4241.664 & -0.1 & & 0.568\\
\enddata
%\tablecomments{}
%can't get tablenotetext to work properly
\end{deluxetable*}

%  blue cross-disperser, blue arm, & Blue 437 nm,  (Image Slicer No.2), Blue arm &

\section{Data Acquisition and Reduction}\label{obs_reduction}
%TO DO
%1. Check if all the values are consistent in the table 
%2. Go through the text with a fine comb, filling in some gaps
%3. Polish the sentences further
A robust abundance analysis of U requires high signal-to-noise (S/N) and high-resolution spectral data, since \utwo\ transition lines are weak and surrounded by blends. We obtained new spectroscopic data for \emhstarShort\ and \vinnistarShort\ using high-resolution spectrographs, Keck/HIRES and Magellan/MIKE, respectively. For \annastar\ and \hillstar, we utilized VLT/UVES archival data.
\par
Following the data reduction and radial-velocity correction, orders of each exposure were normalized using a natural cubic spline function with sigma clipping and the strong lines masked. The normalized orders were co-added and then stitched\footnote{\url{https://github.com/alexji/alexmods/blob/master/alexmods/specutils/continuum.py}} to furnish the final spectrum of each star. We summarize the spectral data properties of all the stars in Table~\ref{tab:observationLog}, including the wavelength range, resolving power, and S/N per pixel of the final spectra.
\subsection{\emhstar}
We observed {\emhstarShort} with Keck/HIRESr \citep{Vogt1994_HIRES} on 2021 March 26 for a total of 6.6h. The observations were broken down as 8 exposures of 1800s, 1 exposure of 1400s, and 1 exposure of 1350s. The observations were taken with the red cross-disperser, using the $5.0\times0\farcs40$ slit and with $1\times1$ binning, which yielded a resolving power of $R \sim 86,\! 600$. We used the blue and the green CCD chip data, which we reduced with \texttt{MAKEE}\footnote{\url{https://sites.astro.caltech.edu/~tb/makee/}} using standard settings. The full wavelength range of the spectra was 3600-6800 {\AA}. We corrected each spectrum for radial velocity by cross-correlating against a high-resolution spectrum of HD\ 122563. We normalized, co-added, and stitched the spectra to furnish the final spectrum with S/N per pixel of 185 at 4050\,\AA. 

\subsection{\vinnistar}
We observed {\vinnistarShort} with Magellan/MIKE \citep{Bernstein2003_MIKE} on 2018 July 8, 2018 July 24, 2018 September 26, and 2018 November 11, for a total of 15.6h. We used the $0\farcs35$ slit and $2\times1$ binning, which yielded a resolving power of $R \sim 83,\! 000$.
%spectra on the blue arm.
%, though in practice the achieved resolving power from arc lines is closer to $65,\! 000$, and the star itself had a resolving power of $55,\! 000$, possibly due to rotation. 
Data for this star already existed \citep{PlaccoJ2038_2017}, but with $2\times2$ binning, which could have undersampled the profiles of the \utwo\ lines, so we reobserved the star. We reduced the spectra from each night, together, using \texttt{CarPy} \citep{Kelson2000_CarPy1stref, Kelson2003_CarPy2ndref} and corrected the radial velocity by cross-correlating against a high-resolution spectrum of HD\ 122563. We normalized, coadded, and stitched the spectra to furnish the final spectrum with S/N per pixel of 175 at 4050{\AA}.

\subsection{\annastar}
We used the data from \cite{FrebelHE1523_2007}, who observed the star with VLT/UVES \citep{Dekker2000_VLT/UVES} in 2005 and 2006, using image slicer No. 2 and $0\farcs45$ slit width to achieve a resolving power of $R \sim 75,\! 000$. The data are publicly available on the ESO raw data archive\footnote{\url{http://archive.eso.org/eso/eso_archive_main.html}}. We used the BLUE 437 nm setting observations from 2006 April 22, 2006 April 23, and 2006 May 19, which amounted to a total exposure time of 3.0h.
%and S/N per pixel of 200 at 4050 {\AA} for the final spectrum. 
The wavelength range of the final spectrum was 3758-4990 {\AA}, which included all the lines of interest. We reduced the data using \texttt{ESOReflex} \citep{Freudling2013_ESOReflex}, with order extraction set to the recommended \texttt{linear} method for data collected with an image slicer. We corrected the radial velocity of each exposure by cross-correlating with a high quality MIKE/Magellan spectrum of {\annastar}. We normalized, co-added, and stitched the spectra to furnish the final spectrum with S/N per pixel of 200 at 4050 {\AA}.

\subsection{\hillstar}
We used data from \cite{HillCS31082_2002}, who observed the star with VLT/UVES \citep{Dekker2000_VLT/UVES} in 2000 using $0\farcs45$ slit-width. The data are publicly available on the ESO raw data archive. We used the BLUE arm 380-510 nm setting observations from 2000 October 17 and 2000 October 19, which totaled 2h of exposure. The resulting resolving power was $R \sim 75,\! 000$. We reduced the data using \texttt{ESOReflex} \citep{Freudling2013_ESOReflex}, with order extraction set to the recommended \texttt{optimal} method. We corrected the radial velocity of each exposure by cross-correlating to a high-quality MIKE/Magellan spectrum of {\annastar}. We normalized, co-added, and stitched the spectra to furnish the final spectrum with S/N of 125 per pixel at 4050 {\AA}.

\begin{deluxetable*}{cccccc}
\tablenum{2}
\tablecaption{Stellar Parameters \label{tab:StellarParams}}
\tablehead{
\colhead{Star} & \colhead{Source} & \colhead{\eTeff} & \colhead{\logg} & \colhead{[Fe/H]} & \colhead{\vmicro} \\
\colhead{} & \colhead{} & \colhead{(K)} & \colhead{(cgs)} & \colhead{} & \colhead{(km s$^{-1}$)}
}

\startdata 
\emhstarShort & This work & $4410 \pm 150$ & $0.61 \pm 0.30$ & $-2.96 \pm 0.14$ & $2.74 \pm 0.20$ \\
& \cite{holmbeckJ0954_2018} & $4340 \pm125$ & $0.41 \pm 0.20$ & $-2.99 \pm 0.10$ & $2.28 \pm 0.20$ \\
& & & & & \\
\vinnistarShort & This work & $4519 \pm 150$ & $0.57 \pm 0.30$ & $-3.12 \pm 0.12$ & $2.26 \pm 0.20$ \\
& \cite{PlaccoJ2038_2017} & $4630 \pm 100$ & $1.20 \pm 0.20$ & $-2.91 \pm 0.10$ & $2.15 \pm 0.20$ \\
& & & & & \\
\annastar & This work & $4607 \pm 150$ & $0.94 \pm 0.30$ & $-2.98 \pm 0.14$ & $2.65 \pm 0.20$\\
& \cite{FrebelHE1523_2007} & $4630 \pm 40$ & $1.00 \pm 0.30$ & $-2.95 \pm 0.2$ & $2.60 \pm 0.30$ \\
& & & & & \\
\hillstar & This work & $4793 \pm 150$ & $1.36 \pm 0.30$ & $-2.94 \pm 0.11$ & $1.68 \pm 0.20$\\
& \cite{HillCS31082_2002} & $4825 \pm 120$ & $1.50 \pm 0.30$ & $-2.9 \pm 0.13$ & $1.80 \pm 0.20$ \\
% & & & & \\
\enddata
%\tablecomments{*We adopt stellar parameters derived by \cite{PlaccoJ2038_2017}.}
\end{deluxetable*}

\section{Stellar Parameters}\label{sec:stellarparams}
We derived the stellar parameters of all the stars spectroscopically. For this purpose, we used \texttt{SMHr}\footnote{We used \url{https://github.com/eholmbeck/smhr-rpa/tree/refactor-scatterplot} forked from \url{https://github.com/andycasey/smhr/tree/refactor-scatterplot}}, the next generation spectroscopic analysis software of \texttt{SMH} \citep{Casey2014_SMH}. \texttt{SMHr} wraps the radiative transfer code, \texttt{MOOG} \citep{Sneden1973_MOOG} and allows the employment of various grid model atmospheres. We used \texttt{MOOG} with the proper treatment of scattering included\footnote{\url{https://github.com/alexji/moog17scat}} \citep{Sobeck2011_MOOG} and employed the \texttt{ATLAS9} grid of 1D plane-parallel model atmospheres computed under the assumption of local thermodynamic equilibrium (LTE) \citep{Castelli&Kurucz2004_ATLAS9}. 
\par
We used equivalent widths (EWs) of {\ion{Fe}{1}} and {\ion{Fe}{2}} lines for stellar parameter determination of all the stars. We measured the EWs within \texttt{SMHr} by fitting Gaussian or Voigt profiles. We obtained the effective temperature ({\eTeff}) by inducing equilibrium in the {\ion{Fe}{1}} line-abundances with respect to the excitation potential of the lines, the surface gravity ({\logg}) by minimizing the difference between the mean {\ion{Fe}{1}} and {\ion{Fe}{2}} abundances, and the microturbulent velocity ({\vmicro}) by inducing an equilibrium in {\ion{Fe}{1}} line-abundances with respect to the reduced EW of the lines. We solved for the stellar parameters, {\eTeff}, {\logg}, and {\vmicro}, simultaneously with multiple iterations and corrected the resulting  best-fit {\eTeff} to the photometric scale described in \citet{Frebel2013_Teff}. Subsequently, we re-derived \logg\ and \vmicro\ as described above for {\eTeff} fixed to this corrected value. We list the resulting stellar parameters of \emhstarShort, \vinnistarShort, \annastar, and \hillstar\ in Table~\ref{tab:StellarParams}. 
\par
The stellar parameters derived in this work agree with those determined previously in the literature within uncertainties for all the stars in our sample, except for \vinnistarShort. The primary disagreement for \vinnistarShort\ is in \logg, where we derived \logg\ $=0.57$, whereas \cite{PlaccoJ2038_2017} derived \logg\ $=1.20$ using the EW technique. Upon further investigation into this discrepancy, we suspect that it mostly originates from different implementations of scattering in \texttt{MOOG}. For homogeneity with the other stars in our sample, we adopt our derived stellar parameters for \vinnistarShort.

\begin{deluxetable*}{cccccc}
\tablenum{3}
\tablecaption{Abundances and Isotopic Ratios of \utwo\ Line Blends \label{tab:Blends}}
\tablehead{
\colhead{} & \colhead{Source} & \colhead{\emhstarShort$^\mathrm{a}$} & \colhead{\vinnistarShort$^\mathrm{b}$} & \colhead{\annastar$^\mathrm{c}$} & \colhead{\hillstar$^\mathrm{d}$}
%\colhead{} & \colhead{} & \colhead{\eps(X)} \\
}
\startdata
\eps(Fe) & This Work & $4.55\pm0.15$ &  $4.41\pm0.12$ & $4.53\pm0.14$  & $4.58\pm0.11$\\
   & Other & $4.51\pm0.12$ & $4.59\pm0.12$ & $4.50\pm0.20$  & $4.60\pm0.13$\\ 
& & & & & \\
\eps(C) & This Work & $4.97\pm0.20$  & $5.11\pm0.20$   & $5.17\pm0.20$  &   $5.75\pm0.20$\\
 & Other & $4.94\pm0.20$  &  $5.08\pm0.20$  & $5.14$  & $5.82\pm0.05$ \\
& & & & & \\
 $\mathrm{^{12}C/^{13}C}$ & This Work & $4.0$ &  $4.6$ & $3.5$ & $19.0$ \\
& Other & \nodata & \nodata & $\sim 3$-$4$ & > 20.0 \\
& & & & & \\
\eps(N) & This Work & $5.82\pm0.20$  &  $5.56\pm0.20$  & $5.88\pm0.20$  &  \nodata \\
 & Other & \nodata & \nodata  &  $5.43$ & <$5.22$ \\
 & & & & & \\
\eps(La) & This Work & $-1.06\pm0.09$  & $-1.06\pm0.05$ & $-0.47\pm0.11$ & $-0.65\pm0.07$\\
    & Other & $-1.15\pm0.10$  & $-0.76\pm0.07$ & $-0.63$ & $-0.60\pm0.04$\\
\hline
\enddata
\tablecomments{Source of other work: $^\mathrm{a}$\cite{holmbeckJ0954_2018},$^\mathrm{b}$\cite{PlaccoJ2038_2017},$^\mathrm{c}$\cite{FrebelHE1523_2007}, $^\mathrm{d}$\cite{HillCS31082_2002}.}
\end{deluxetable*}

\section{Chemical-Abundance Analysis}\label{sec:chemAbund}
We derived chemical abundances for all the stars using EWs and spectral synthesis in \texttt{SMHr}. Though abundances for these stars have been previously reported in the literature, we re-derived abundances of the relevant elements for a homogeneous and consistent analysis. This enabled us to robustly constrain the transitions directly blended with the \utwo\ lines, as well as those neighboring the \utwo\ lines, which could affect the local continuum placement. We derived abundances of most light elements, including  Na, Mg, Al,  Si, Ca, Ti, Cr, Fe, and Zn, using the EW method. We derived abundances of the remaining light elements and the neutron-capture (n-cap) elements, including C, N, V, Mn, Sr, Y, Zr, Ba, La, Ce, Pr, Nd, Sm, Eu, Gd, Dy, Tm, Er, Th, and U, with spectral synthesis of $\pm5$\,\AA\ regions around the transition lines. For the abundance determination of the light and n-cap elements, we used a subset of the transition list compiled by \cite{Roederer2018_HD222925}. For abundance analysis with spectral synthesis, we generated the atomic-parameters linelists with \texttt{linemake}\footnote{\url{https://github.com/vmplacco/linemake}} \citep{linemake2021}, which included the transition wavelengths ($\lambda$), excitation potentials ($\chi$), oscillator strengths (\loggf), and hyperfine structure of the transition lines. We used the updated atomic parameters of CH transitions from \cite{Masseron2014_fixed}\footnote{\url{https://github.com/alexji/linemake}} and $r$-process isotopic ratios from \cite{Sneden2008rev_isotopicRatios} for the spectral synthesis of Ba, Eu, Nd, Sm, Yb, and Pb. We further detail the abundance determination of the \utwo\ line blends, U, Th, and Eu in sections \ref{subsec:blends}, \ref{subsec:U}, \ref{subsec:Th}, and \ref{subsec:Eu}, respectively. We describe uncertainty analysis of the derived abundances in section \ref{subsec:uncertainty}.

\begin{deluxetable}{ccccc}
\tablenum{4}
\tablecaption{Atomic Parameters of \utwo\ Transition Lines. \label{tab:AtomicData}
}
\tablehead{
\colhead{Species} &  \colhead{$\lambda$ (\AA)}  & \colhead{$\chi$ (eV)} & \colhead{\loggf} & \colhead{\% Uncertainty} \\
\colhead{} & & & & \colhead{in $gf$}
}
\startdata
\utwo\ & 3859.57  & 0.036 & $-0.067$ & 12  \\
\utwo\ & 4050.04 &  0.000 & $-0.706$ &  7\\
\utwo\ & 4090.13  & 0.217 & $-0.184$ &  13 \\
\enddata
\tablecomments{\loggf\ values and \% uncertainty on $gf$ values taken from Table 2 of \cite{Nilsson2002}. Excitation potential taken from \texttt{linemake}.}
\end{deluxetable}

\begin{figure*}
    \centering
    \includegraphics[width = 0.9\textwidth]{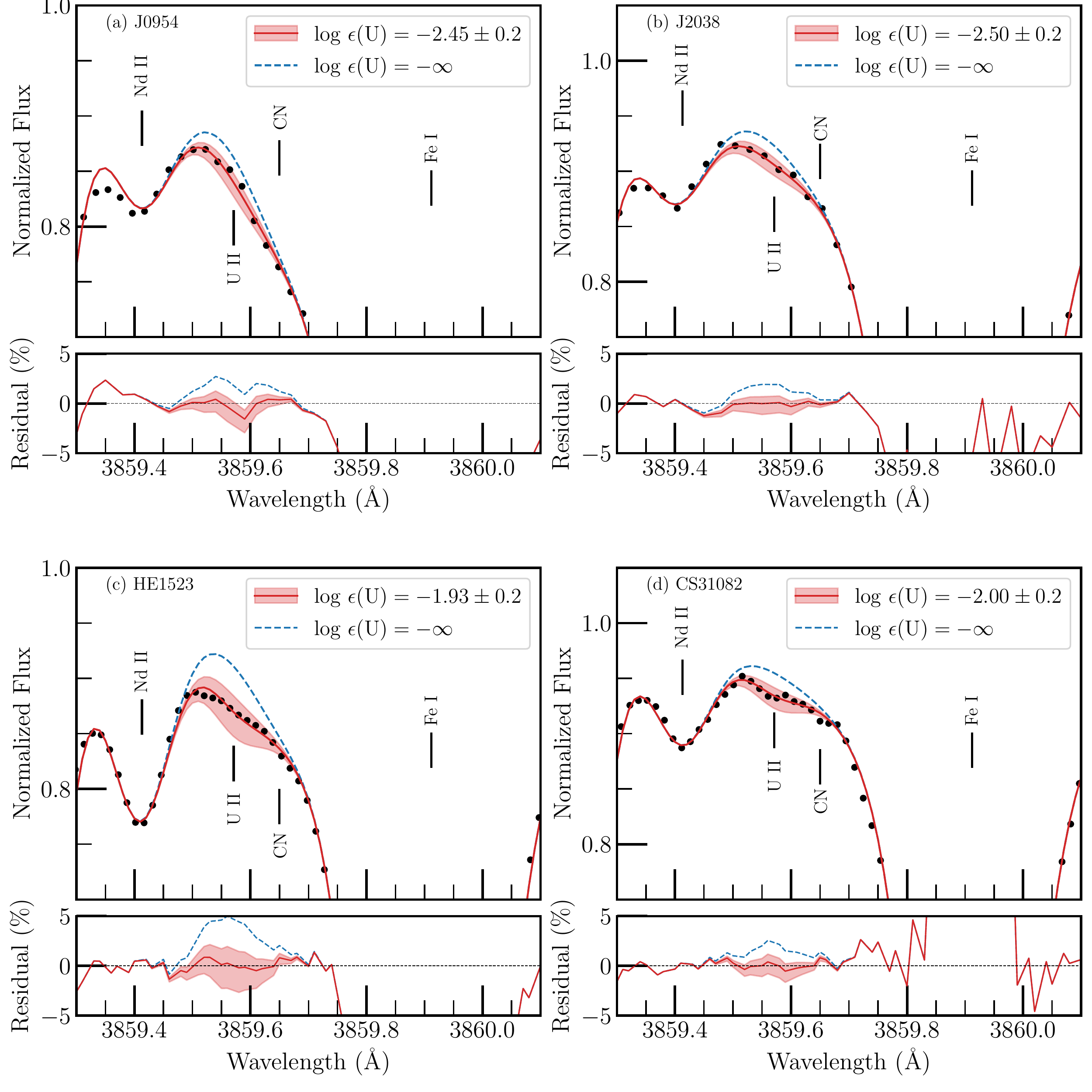}
    \caption{Spectral synthesis of the \utwo\ line at $\lambda3859.57$\,\AA\ for \emhstarShort, \vinnistarShort, \annastar, and \hillstar. The red-solid line traces the best-fit synthetic model to the observed data in black points. The red-shaded region depicts abundance variation within $\pm\ 0.2$ dex of the best-fit U abundance. The blue-dashed line traces the synthetic model with no U. Important neighboring transition lines are labeled. The corresponding residuals between the observed data and the synthetic models are also shown.}
    \label{fig:3859ZoomIn}
\end{figure*}

\begin{figure*}
    \centering
    \includegraphics[width = 0.9\textwidth]{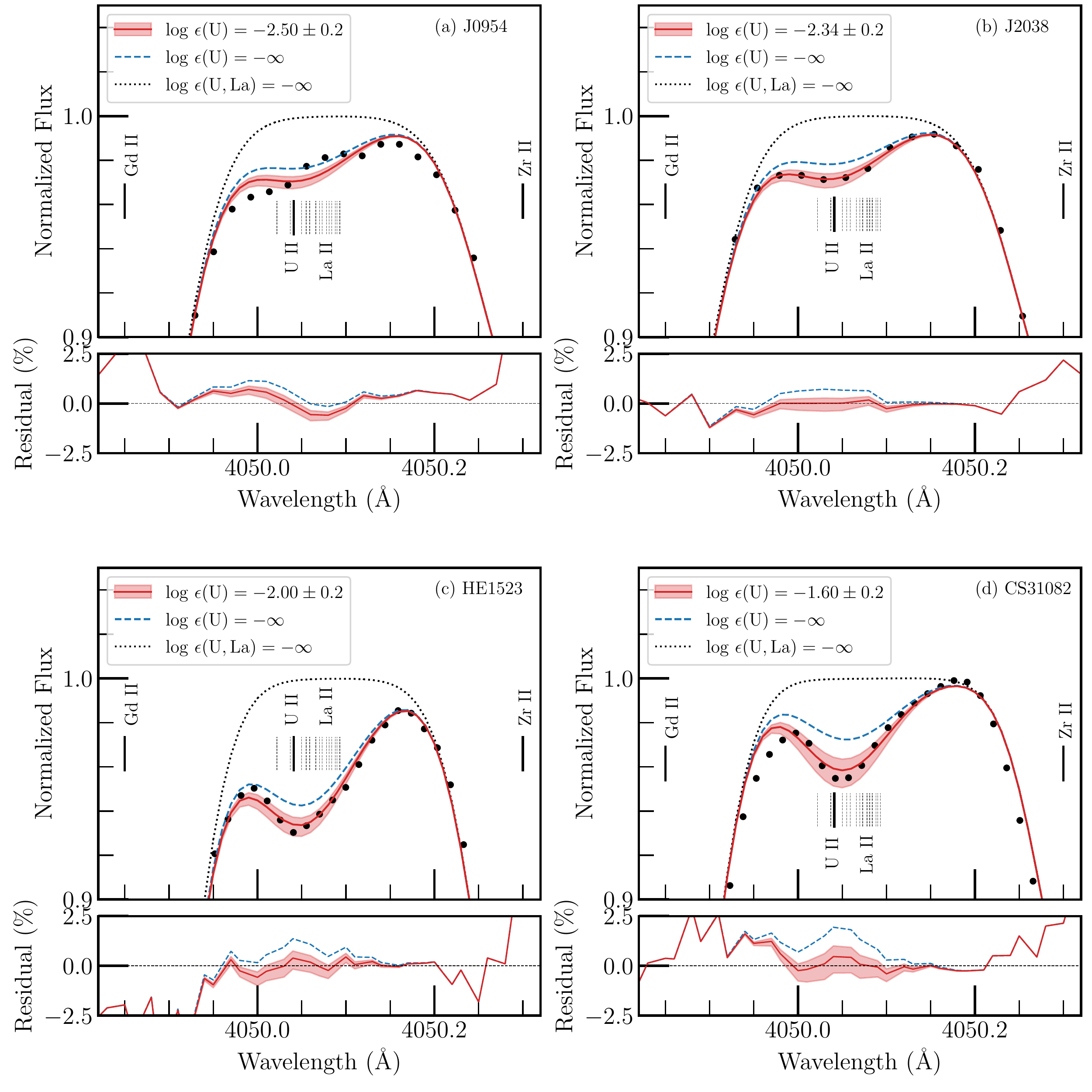}
    \caption{Spectral synthesis of the \utwo\ line at $\lambda4050.04$\,\AA\ for \emhstarShort, \vinnistarShort, \annastar, and \hillstar. The red-solid line traces the best-fit synthetic model to the observed data in black points. The red-shaded region depicts the abundance variation within $\pm\ 0.2$ dex of the best-fit U abundance. The blue-dashed line traces the synthetic model with no U, and the black-dotted line traces the synthetic model with no U and no La. Important neighboring transition lines are labeled. The corresponding residuals between the observed data and synthetic models are also shown.}
    \label{fig:4050EditedZoomIn}
\end{figure*}
\begin{figure*}
    \centering
    \includegraphics[width = 0.9\textwidth]{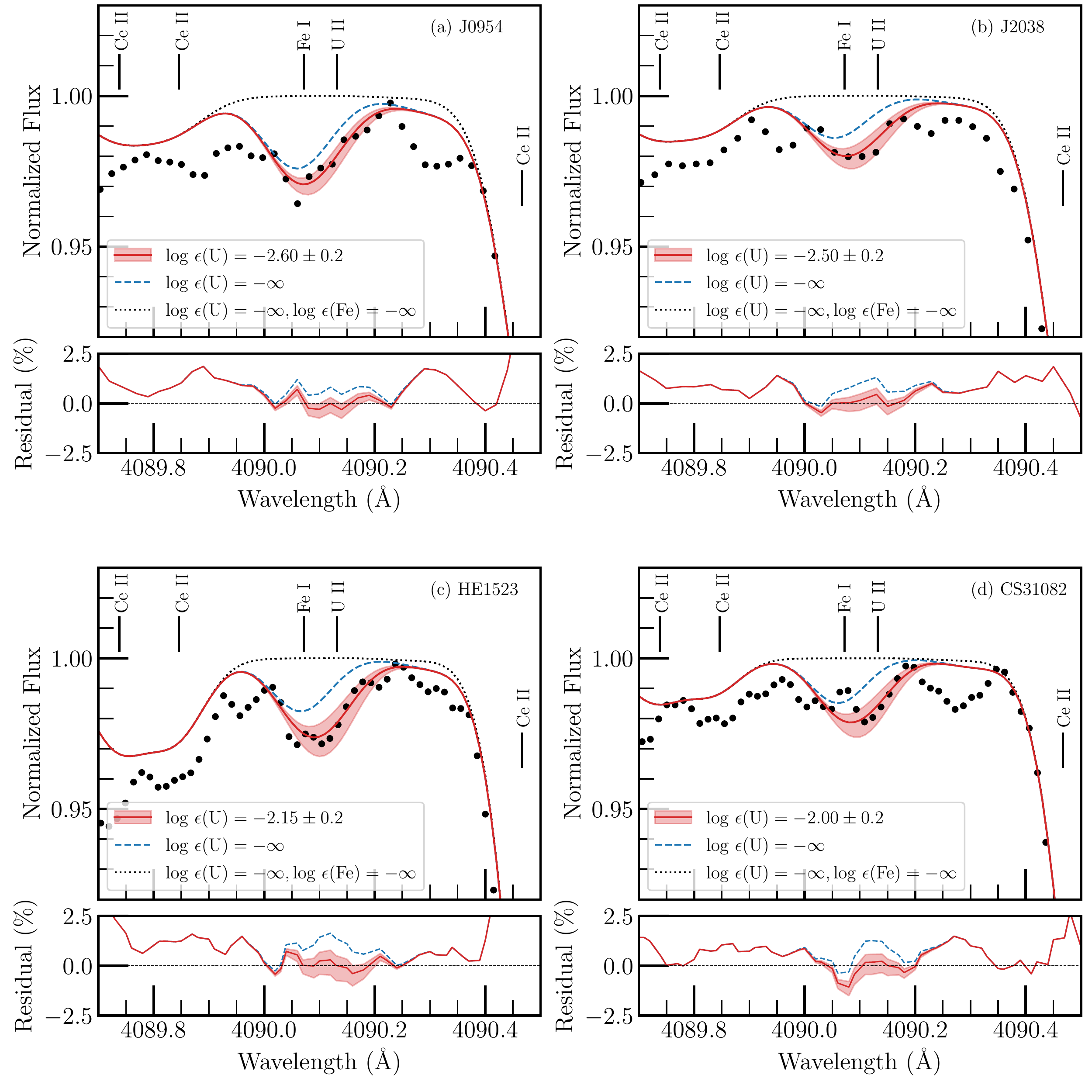}
    \caption{Spectral synthesis of the \utwo\ line at $\lambda4090.13$\,\AA\ for \emhstarShort, \vinnistarShort, \annastar, and \hillstar. The red-solid line traces the best-fit synthetic model to the observed data in black points. The red-shaded region depicts the abundance variation within $\pm\ 0.2$ dex of the best-fit U abundance. The blue-dashed line traces the synthetic model with no U, and the black-dotted line traces the synthetic model with no U and no Fe. Important neighboring transition lines are labeled. The corresponding residuals between the synthetic models and the observed data are also shown.}
    \label{fig:4090ZoomIn}
\end{figure*}

\begin{figure}
    \centering
    \includegraphics[width = 0.47\textwidth]{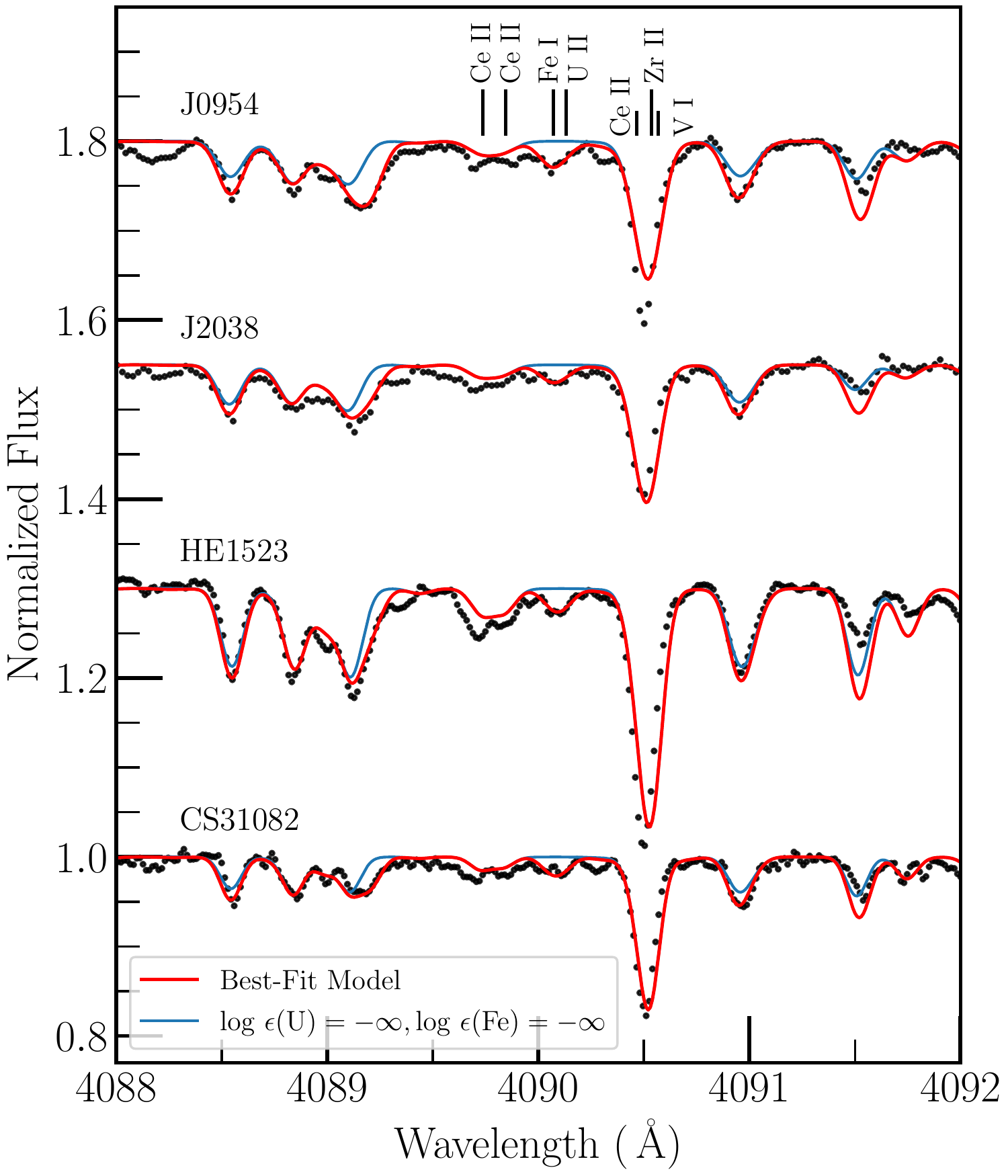}
    \caption{Spectral synthesis around the $\lambda4090.13$\,\AA\ \utwo\ line region for \emhstarShort, \vinnistarShort, \annastar, and \hillstar. The normalized flux of the stars are scaled for illustration. The red-solid line traces the best-fit synthetic model to the observed data in black points. The blue-solid line traces the synthetic model with no U and no Fe, depicting the continuum at the \utwo\ transition. Important neighboring transition lines are labeled. This zoomed-out plot depicts the best placement of the local continuum of the spectral synthesis models for our sample stars.}
    \label{fig:4090ZoomOut}
\end{figure}
\newpage
\subsection{Blends: Fe, C, N, and La}\label{subsec:blends}
We took special care to constrain the abundances of elements that have transitions blended with the weak \utwo\ lines. We identified that transitions of \ion{Fe}{1}, CH, CN, and \ion{La}{2} are blended with the \utwo\ lines investigated in this work. We obtained the Fe abundance using EW measurements of a subset of acceptable \ion{Fe}{1} lines listed in \cite{Roederer2018_HD222925}, for each sample star. We estimated the uncertainty on the mean Fe abundance as the standard deviation in the abundances of the chosen \ion{Fe}{1} lines. We determined the C abundance by fitting the $\lambda$4313\,\AA\ $G$-band of CH. Based on the quality of the data and the synthetic spectrum fits, we set a fiducial uncertainty estimate of $\pm0.2$ dex on the C abundance of all the sample stars. With the C abundance fixed, we determined the isotopic ratio of $\mathrm{^{12}C/^{13}C}$ by fitting the $^{13}$CH feature at $\lambda$4217\,\AA. We determined the N abundance by fitting the $\lambda$3876\,\AA\ CN molecular band for all our sample stars, except for \hillstar, for which we could not derive a reliable N abundance. For the N abundance of \emhstarShort, \vinnistar, and \annastar, we estimated an uncertainty of $\pm0.2$ dex, based on the spectral synthesis fits. For each sample star, we determined the La abundance with the spectral synthesis of a subset of blend-free and acceptable \ion{La}{2} transitions listed in \cite{Roederer2018_HD222925}. We estimated the uncertainty on the mean La abundance as the standard deviation in the La abundances of the chosen \ion{La}{2} lines. We list the Fe, C, N, and La abundances and the $\mathrm{^{12}C/^{13}C}$ isotopic ratio, determined for each star, in Table~\ref{tab:Blends}, along with their corresponding values from previous literature studies. The abundances determined in this work are in agreement with the values from the literature, within uncertainties. The exception to this case is our derived La abundance for \vinnistarShort, which disagrees with the abundance derived by \cite{PlaccoJ2038_2017}. However, this discrepancy is simply attributed to the difference in our adopted stellar parameters (see section \ref{sec:stellarparams}).

\subsection{Uranium}\label{subsec:U}
So far in the literature, U abundances of RPE stars have been primarily determined using a single \utwo\ line at $\lambda$3859\,\AA\footnote{An exception to this case is \cite{Roederer2018_HD222925}, who also used the \utwo\ line at $\lambda$4241\,\AA\ to place an upper limit on the U abundance of an RPE star, HD 222925.}. Moreover, U abundance analyses have been carried out by different studies for individual stars, with each study varying in the employed method and atomic data.  
\par
In this study, we performed, for the first time, a homogeneous analysis to determine U abundances of four highly RPE stars using three \utwo\ lines at $\lambda$3859\,\AA, $\lambda$4050\,\AA, and $\lambda$4090\,\AA. We generated the linelist for the spectral synthesis of the \utwo\ line-regions with \texttt{linemake}. We used the \loggf\ measurements of the \utwo\ lines from \cite{Nilsson2002}, who measured them with high accuracy by combining their branching fraction calculations with the radiative lifetime measurements of 6 \utwo\ levels from \cite{Lundberg2001}. We list the atomic parameters employed for the three \utwo\ transitions in Table~\ref{tab:AtomicData}.  
\par
We determined the final U abundance of each star as the weighted-average of the U abundances from the three \utwo\ lines i.e., \epsU$ \sum_{i} (w_{i}\log \epsilon_{i})/\sum_{i} w_{i}$, where $\log \epsilon_{i}$ is the U abundance from line $i$ and $w_{i} = 1/\Delta\mathrm{(stat)}_{i}^{2}$ for line $i$ \citep{Ji2019, McWilliam1995}. We detail the method we used to estimate $\Delta\mathrm{(stat)_i}$, the statistical uncertainty on $\log \epsilon_{i}$, in Section~\ref{subsec:uncertainty}. For the total uncertainty on the average-weighted U abundances, we accounted for systematic uncertainties (from stellar parameters and blends) and statistical uncertainties (from \loggf\ measurement and continuum placement). We discuss this further in Section~\ref{subsec:uncertainty}. We list the final weighted-average U abundance with the associated total uncertainty for all the sample stars in Table~\ref{tab:Abundances}. 
%We computed the standard-error, $\sigma_{w}$, of the weighted-average U abundance as $1/\sigma_{w}^{2} = \sum_{i} w_{i}$ for all the sample stars. For the total uncertainty, $\Delta$(total), on the weighted-average U abundance, we also accounted for $\Delta$(sys), the systematic uncertainty from the stellar parameters (see section~\ref{subsec:uncertainty}). We computed $\Delta$(total) as the quadrature sum of $\sigma_{w}$ and $\Delta$(sys). We list the final weighted-average U abundance with the associated total uncertainty for all the sample stars in Table~\ref{tab:Abundances}. 
%We list $\log \epsilon_{i   }$ and its associated $\sigma_{i}$ for each \utwo\ line in Table~\ref{tab:Abundances} for all the sample stars.
\subsubsection{The $\lambda$3859\,\AA\ \ion{U}{2} Line}
While the $\lambda3859.57$\,\AA\ line is the strongest \utwo\ line discernible in the spectra of stars, blends from other transitions makes its spectral synthesis quite difficult. This line is situated in the wing of a strong \ion{Fe}{1} line at $\lambda$3859.91\,\AA\ and is further blended with a CN feature at $\lambda$3859.65\,\AA, which also resides in the wing of the \ion{Fe}{1} line. Therefore, it is essential to constrain the wing of the \ion{Fe}{1} line as well as the CN feature for a reliable U abundance determination.
\par
To fit the wings of the strong \ion{Fe}{1} line, we used the Uns\"{o}ld approximation \citep{unsold} multiplied by a factor of $-8.77$ for the Van der Waals hydrogen collision-damping coefficient of the \ion{Fe}{1} line, for all the sample stars. Furthermore, we adjusted the derived \ion{Fe}{1} abundances of the stars by $-0.15$, $-0.05$, $+0.02$, and $+0.03$ dex for \emhstarShort, \vinnistarShort, \annastar, and \hillstar, respectively. For most of the stars, the adjustment made to the derived \ion{Fe}{1} abundance of the star is small and lies within the uncertainty on the \ion{Fe}{1} abundance. 
\par
To fit the CN feature, we adjusted the derived N abundance of the stars by $+0.0$, $+0.04$, and $+0.12$ dex for \emhstarShort, \vinnistarShort, and \annastar, respectively. We find these adjustments acceptable, since they are within the $\pm0.2$ dex uncertainty on the derived N abundances. For \hillstar, we used \eps(N) $= 5.22$, which is the upper limit placed on the N abundance by \cite{HillCS31082_2002}, as we could not derive a reliable N abundance for the star. We used the derived C abundance of the stars without any adjustments.
\par
For the purpose of a better fit to the neighboring line features, we blue-shifted the transition wavelength of \ion{Nd}{2} line at $3859.42$\,\AA\ and \ion{Fe}{1} line at $3859.21$\,\AA\ by $0.07$\,\AA. To fit the \ion{Nd}{2} line, we adjusted the derived \ion{Nd}{2} abundance of the star by $-0.12$, $+0.0$, $+0.04$, and $-0.05$ dex for \emhstarShort, \vinnistarShort, \annastar, and \hillstar, respectively. The adjustment is small for most stars and within the uncertainty of the Nd abundance. 
\par
We determined \eps(U)$_{3859} = -2.45$, $-2.50$, $-1.93$, and $-2.0$ for \emhstarShort, \vinnistarShort, \annastar, and \hillstar, respectively. In Figure~\ref{fig:3859ZoomIn}, we show the resulting best-fit spectral synthesis model for the observed data of each star, along with the residuals between the model and the data. We also depict the $\pm0.2$ dex abundance variation from the derived $\lambda$3859 U abundance in red-shaded region, for all the sample stars.

\subsubsection{The 4050\,\AA\ \ion{U}{2}}
The $\lambda4050.04$\,\AA\ \utwo\ line is blended with a \ion{La}{2} line at $\lambda4050.07$\,\AA, which needs to be constrained well for U abundance determination. \texttt{linemake} obtains \loggf\ $=0.11$ for the \ion{La}{2} line from \cite{CorlissBozman1962_Laloggf}, but we used an updated \loggf\ $=0.428$, as measured by \cite{Bord96}. We substantiated this choice with spectral synthesis of the \ion{La}{2} line in RPE stars with minimal U contamination, which showed that the \cite{Bord96} value provides a better fit to the observed spectra of these stars. Additionally, we blue-shifted the transition wavelength of the \ion{La}{2} line by $0.02$\,\AA\ to enable a better fit to the observed data. We also account for the hyperfine splitting (HFS) structure of this \ion{La}{2} line, as described in Appendix~\ref{Appendix_la4050hfs}. We applied the described prescription for the \ion{La}{2} line uniformly across all the sample stars to determine their U abundances. We employed the derived La abundance of each star in the spectral synthesis without any adjustment.
\par
We determined \eps(U)$_{4050} = -2.50, -2.34, -2.00$, and $-1.60$ for \emhstarShort, \vinnistarShort, \annastar, and \hillstar, respectively. We show the corresponding best-fit spectral synthesis models for all the stars in Figure~\ref{fig:4050EditedZoomIn}, along with the resulting residuals between the model and the observed data. We also depict $\pm0.2$ dex variations in the $\lambda4050$ U abundance with a red-shaded region for all the stars. We generally find a good fit to the $\lambda4050$\,\AA\ spectral region, as seen in Figure~\ref{fig:4050EditedZoomIn}. We note an over-estimation of the synthetic model flux around $\lambda4049.95$\,\AA\ for \emhstarShort\ and \hillstar. This could possibly indicate an unidentified line between the \ion{Gd}{2} and \utwo\ lines that has manifested itself more strongly in \emhstarShort\ and \hillstar, as compared to in \vinnistarShort\ and \annastar. Alternatively, the abundance of the \ion{La}{2} HFS structure may not be well represented by the mean La abundances determined for the stars. 
\newline
\subsubsection{The $\lambda4090$\,\AA\ \utwo\ line}
The $\lambda4090.13$\,\AA\ \ion{U}{2} line is blended with one weak \ion{Fe}{1} line at $\lambda4090.07$\,\AA. We derived the $\lambda4090$ U abundance of all the sample stars with spectral synthesis, specifically, by fitting the \utwo\ line to the red-ward wing of the Fe-U absorption feature. We determined  \eps(U)$_{4090} = -2.60, -2.50, -2.15, \rm{and} -2.00$ for \emhstarShort, \vinnistarShort, \annastar, and \hillstar, respectively. We show the corresponding best-fit spectral synthesis models for all the stars in Figure~\ref{fig:4090ZoomIn}, along with residuals between the models and the observed data. We also depict $\pm0.2$ variation in the best-fit $\lambda4050$ U abundance with red-shaded region for all the sample stars. We note that the two absorption features blue-ward and red-ward of the \ion{U}{2} line are currently unidentified in the \texttt{linemake} linelist. We tested the effect of these unidentified features by adding ``fabricated'' lines to mimic them and found that they have minimal-to-no effect on the U abundance determination. In Figure \ref{fig:4090ZoomOut}, we also show the best-fit spectral synthesis for a wider wavelength window of this line region, for all the stars. This figure depicts that even though the immediately neighboring  lines of the $\lambda4090.13$\,\AA\ \ion{U}{2} line are unidentified, we found an optimal continuum placement for all the stars using other spectral regions. 

\begin{figure*}[ht!]
    \centering
    \includegraphics[width = 0.9\textwidth]{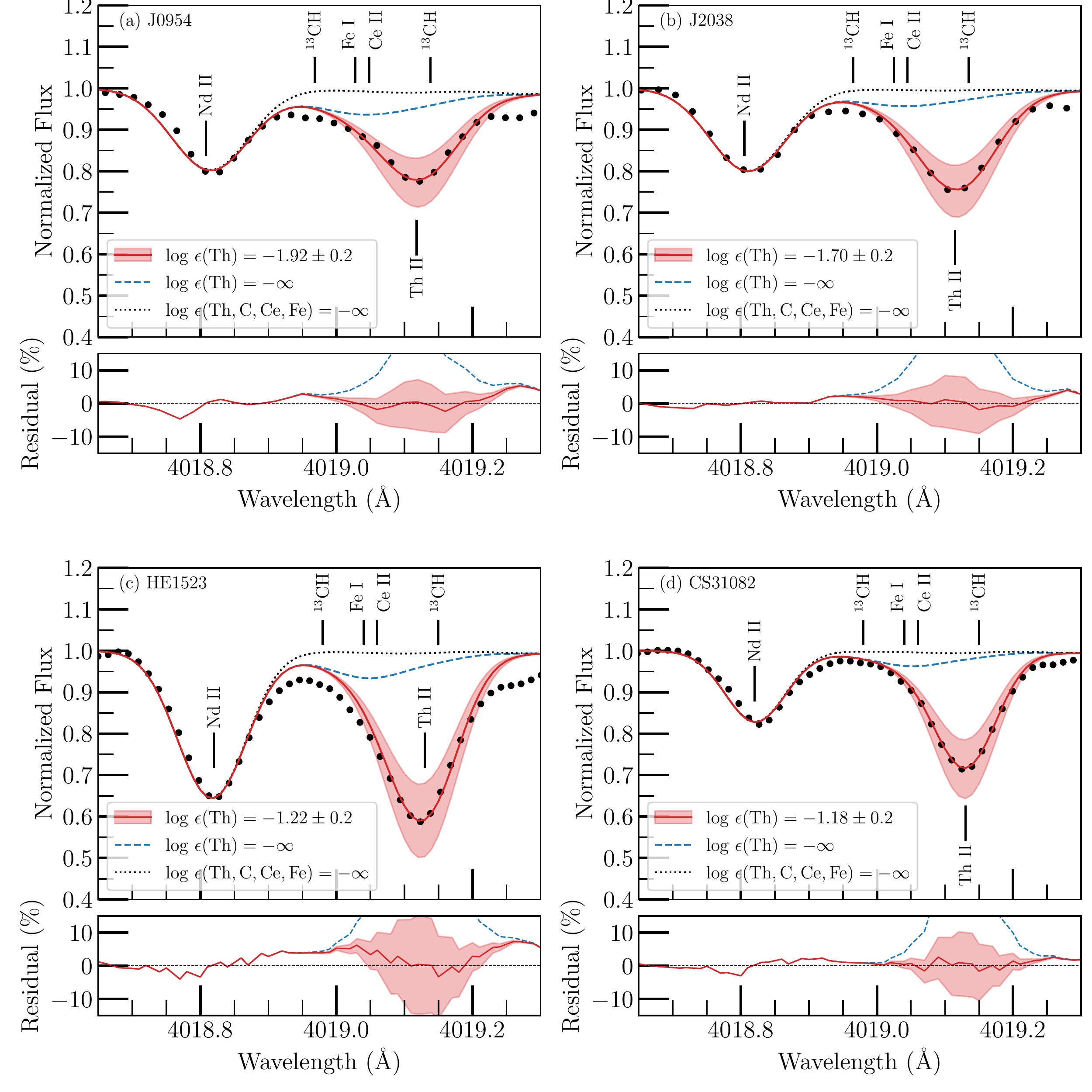}
    \caption{Spectral synthesis of the \thtwo\ line at $\lambda4019.13$\,\AA\ for \emhstarShort, \vinnistarShort, \annastar, and \hillstar. The red-solid line traces the best-fit synthetic model to the observed data in black points. The blue-dashed line traces the synthetic model with no Th, and the black-dotted line traces the synthetic model with no Th, Fe, C, and Ce. The red-shaded region depicts abundance variations within $\pm0.2$ dex. Important neighboring transition lines are labeled. The corresponding residuals between the synthetic models and the observed data are also shown.}
    \label{fig:Th4019}
\end{figure*}
\begin{deluxetable*}{cccccc}
\tablenum{5}
\tablecaption{U, Th, and Eu Abundances and Nucleocosmochronometric Ages.* \label{tab:Abundances}}
\tablehead{
\colhead{Source}& \colhead{\eps(X)} & \colhead{\emhstarShort$^\mathrm{a}$} & \colhead{\vinnistarShort$^\mathrm{b}$} & \colhead{\annastar$^\mathrm{c}$} & \colhead{\hillstar$^\mathrm{d}$}
}
\startdata 
Other Work & \eps(U)$_{3859}$ & $-2.13 \pm\ 0.20$  &  $-2.14 \pm\ 0.20$ &  $-2.06 \pm\ 0.12$ &  $-1.92 \pm\ 0.17$ \\
This Work & \eps(U)$_{3859}$ &  $-2.45 \pm\ 0.30$  & $-2.50 \pm\ 0.26$ &  $-1.93 \pm\ 0.18$ & $-2.00 \pm\ 0.22$ \\
 & \eps(U)$_{4050}$&            $-2.50 \pm\ 0.33$  & $-2.34 \pm\ 0.30$ &  $-2.00 \pm\ 0.48$ &  $-1.60 \pm\ 0.21$ \\
 & \eps(U)$_{4090}$ &           $-2.60 \pm\ 0.30$ &  $-2.50 \pm\ 0.24$ &  $-2.15 \pm\ 0.28$ &  $-2.00 \pm\ 0.25$ \\
 &  \eps(U) &                   $-2.50 \pm\ 0.29$  &  $-2.47\pm0.21$ & $-1.96 \pm\ 0.25$ &  $-1.87 \pm\ 0.19$ \\
& &  &  & &  \\
Other Work & \eps(Th) &         $-1.13\pm0.10$ & $-1.24\pm0.10$ & $-1.2\pm0.05$ & $-0.98\pm0.13$ \\
This Work & \eps(Th)$_{4019}$ & $-1.92\pm0.09$ & $-1.70\pm0.05$  & $-1.22\pm0.11$  & $-1.18\pm0.04$ \\ 
& \eps(Th)$_{4086}$ &           $-1.70\pm0.09$  & $-1.63\pm0.05$  & $-0.95\pm0.11$  & $-1.09\pm0.04$ \\    
& \eps(Th)$_{4095}$ &           $-1.76\pm0.09$  & $-1.57\pm0.05$  & $-1.01\pm0.11$ & $-1.10\pm0.04$ \\
& \eps(Th) &                    $-1.79\pm0.18$ & $-1.63\pm0.21$ & $-1.06\pm0.19$ & $-1.12\pm0.16$ \\
& & & & & \\
Other Work & \eps(Eu) &         $-1.19\pm0.10$ & $-0.75\pm0.10$  &  $-0.62\pm0.05$ & $-0.76\pm0.13$ \\
This Work & \eps(Eu) &          $-1.16\pm0.12$ & $-1.16\pm0.13$  & $-0.53\pm0.08$  & $-0.81\pm0.12$ \\
& & & & & \\
& & & & & \\
\hline
& Chronometer & \emhstarShort\ & \vinnistarShort & \annastar\ & \hillstar\ \\
\hline
Age (Gyr) & U/Th &                      $11.1\pm6.4$ &         $13.5\pm4.8\ $ &                $16.6\pm5.1\ $ &            $11.1\pm4.0\ $\\
($\mathrm{\pm sys\pm stat \pm PR}$) & & ($\pm5.7\pm1.9\pm2.2$) & $(\pm3.8\pm2.1\pm2.2)$ & $(\pm4.2\pm2.0\pm2.2)$ & $(\pm2.8\pm1.8\pm2.2)$ \\
& & & & & \\
Age (Gyr) & U/Eu &                         $12.0\pm4.3$ &         $11.3\pm3.1\ $ &              $14.2\pm3.8\ $ &          $7.3\pm2.6\ $\\
($\mathrm{\pm sys\pm stat \pm PR}$)& & ($\pm3.9\pm1.0\pm1.6$) & $(\pm2.2\pm1.4\pm1.6)$ &  $(\pm3.3\pm1.0\pm1.6)$ &  $(\pm1.6\pm1.3\pm1.6)$ \\
\enddata
\tablecomments{*Nucleocosmochronometric ages listed are obtained in this work. See text for details on uncertainty estimation for the elemental abundances and stellar ages. Source of other work: $^\mathrm{a}$\cite{holmbeckJ0954_2018},$^\mathrm{b}$\cite{PlaccoJ2038_2017},$^\mathrm{c}$\cite{FrebelHE1523_2007}, $^\mathrm{d}$\cite{HillCS31082_2002}. 
}
\end{deluxetable*}
\subsection{Thorium}\label{subsec:Th}
We determined Th abundances for all of the sample stars using \ion{Th}{2} lines at $\lambda4019.13$\,\AA, $\lambda4086.52$\,\AA, and $\lambda4094.75$\,\AA. We generated the linelists for spectral synthesis with \texttt{linemake}, using \loggf\ values from \cite{Nilssson2002_ThII}.
\par
The \thtwo\ line at $\lambda4019.13$\,\AA\ is the strongest Th line detectable in the optical spectra of stars. It is blended with a \ion{Ce}{2} line at $\lambda4019.06$\,\AA, a \ion{Fe}{1} line at $\lambda4019.04$\,\AA\, and $^{13}$CH lines at $\lambda4018.98$\,\AA\ and $\lambda4019.15$\,\AA\ (see Figure~\ref{fig:Th4019}). For the spectral synthesis of this region, we employed the abundances of the blends without any adjustments. We determined \eps(Th)$_{4019} = -1.92, -1.70, -1.22, \rm{and} -1.18$ for \emhstarShort, \vinnistarShort, \annastar, and \hillstar, respectively. The corresponding best-fit spectral synthesis model is shown in Figure~\ref{fig:Th4019} for each star, along with the residual between the mode and the observed data. We note that the synthetic spectrum is overestimated around the $\lambda4018.98$\,\AA\ and $\lambda4019.25$\,\AA\ regions for all the sample stars. This suggests a need to revisit the atomic parameters of the lines in this spectral region and perhaps identify unknown transitions. Nevertheless, we expect minimal effect of these wing-features on the Th abundance, which was robustly determined by constraining the fit of the synthetic spectrum to the core of the absorption feature.
\par
The \thtwo\ line at $\lambda4086.52$\,\AA\ is situated next to a \ion{La}{2} line and partly blended with a \ion{Ce}{2} line. With spectral synthesis of this line-region, we determined \eps(Th)$_{4086} = -1.70, -1.63, -0.95, \rm{and} -1.09$ for \emhstarShort, \vinnistarShort, \annastar, and \hillstar. For the spectral synthesis of \hillstar, we adjusted the derived Ce abundance of the star by $-0.05$ dex.
\par
The \thtwo\ line at $\lambda4094.75$\,\AA\ is blended with a CH line and partly blended with an \ion{Er}{2} line. For the purpose of a good spectral synthesis fit to the region, we allowed an adjustment of the Er abundance within $\pm0.2$ dex of the derived Er stellar abundance for all of the sample stars. Since the \ion{Er}{2} line is blended with only a section of the blue-ward wing of the \thtwo\ line, any adjustment of the Er abundance had minimal effect on the synthetic-spectrum fit to the core of the \ion{Th}{2} line. Subsequently, we determined \eps(Th)$_{4095} = -1.76, -1.57, -1.01, \rm{and} -1.10$ for \emhstarShort, \vinnistarShort, \annastar, and \hillstar, respectively.
\par
The final Th abundance of each sample star was obtained as the mean of the $\lambda4019.13$, $\lambda4086.52$, and $\lambda4094.75$ Th abundances. We determined mean Th abundance as \eps(Th)$= -1.79$, $-1.31$, $-1.06$, and $-1.12$ for \emhstarShort, \vinnistarShort, \annastar, and \hillstar, respectively. We list the Th abundances obtained for each line and the corresponding mean Th abundance for each star in Table~\ref{tab:Abundances}, along with the uncertainty estimates. Table~\ref{tab:Abundances} also lists the Th abundances determined in previous literature studies for comparison. For \emhstarShort, \annastar, and \hillstar, we obtain good agreement with the Th abundances published in the literature. For \vinnistarShort, we note some discrepancy, which we attribute to the difference in the adopted stellar parameters (see section \ref{sec:stellarparams}).
\newpage
\subsection{Europium}\label{subsec:Eu}
We determined the Eu abundance of each sample star using \ion{Eu}{2} lines at $\lambda4219$\,\AA\, $\lambda4205$\,\AA\, and $\lambda4435$\,\AA. We determined the mean \eps(Eu)$ = -1.16, -1.16, -0.53$, and $-0.81$ for \emhstarShort, \vinnistarShort, \annastar, and \hillstar, respectively. We list the mean Eu abundance, along with the uncertainty estimates and Eu abundance estimates from previous literature studies in Table~\ref{tab:Abundances}. For \emhstarShort, \annastar, and \hillstar, we find our derived Eu abundances to be in good agreement with the literature estimates within uncertainties. For \vinnistarShort, we note a discrepancy in the abundances, which we attribute to the difference in the adopted stellar parameters.
\newpage
\subsection{Uncertainty Analysis}\label{subsec:uncertainty}
For the U abundances of the sample stars, we homogeneously accounted for various sources of systematic ($\Delta$(sys)) and statistical ($\Delta$(stat)) uncertainties. For $\Delta$(sys), we considered the uncertainties on the stellar parameters (\eTeff, \logg, and \vmicro) and the abundances of the blending elements. We list the individual systematic uncertainty components, $\Delta$\eTeff, $\Delta$\logg, $\Delta$\vmicro, and $\Delta$(blend) in Table~\ref{tab:uncertainties}. For $\Delta$(stat), we considered the uncertainties on the \loggf\ values ($\Delta$(loggf)) and the continuum placement of the synthetic spectra (($\Delta$(cont)), which we also list in Table~\ref{tab:uncertainties}. For each \utwo\ line, we estimated the individual uncertainty components, $\Delta$\eTeff, $\Delta$\logg, $\Delta$\vmicro, $\Delta$(blend), $\Delta$(loggf), and $\Delta$(cont). 
\par
To estimate U abundance uncertainties from stellar parameters, we independently changed each stellar parameter by its uncertainty. We thus changed \eTeff\ by $+150$ K, \logg\ by $+0.3$ dex, and \vmicro\ by $0.2$ km/s, for all the stars. For every stellar parameter, we re-derived the abundances of key elements and re-synthesized the \utwo\ lines. We report the resulting change in the U abundances as $\Delta$\eTeff, $\Delta$\logg, and $\Delta$\vmicro, for the respective stellar parameter.
\par
We estimated $\Delta$(blend) by changing the abundance of the blending element (e.g., Fe and La) by $\pm1\sigma$ and re-deriving the U abundance. Here $\sigma$ is the standard deviation in the abundance of the blending element, which we have also adopted as the uncertainty on the mean abundance of the respective blending element (see section~\ref{subsec:blends}). We further limited the change in the abundance of the blending to ensure that the new synthetic spectrum flux was within the S/N of the observed spectrum. We considered the following blending elements: Fe for the $\lambda3859$\,\AA\ and $\lambda4090$\,\AA\ \utwo\, lines and La for the $\lambda4050$\,\AA\ \utwo\ line. While the $\lambda3859$\,\AA\ \utwo\ line is also blended with a CN feature, we find that the U abundance determination is most sensitive to the Fe abundance. 
\par
We estimated $\Delta$(\loggf) through varying the \loggf\ values by the measurement uncertaintues listed in \cite{Nilsson2002} and re-deriving the U abundances. We estimated $\Delta$(cont) for each \utwo\ line by changing the local continuum placement in the spectral synthesis of the \utwo\ line by $\pm0.5\%$ and then re-deriving the U abundance. 
\par
We determined the total uncertainty ($\Delta$(total)) on the U abundance of each \utwo\ line as quadrature sum of the \utwo\ line's systematic and statistical uncertainties. In turn, we determined $\Delta$(sys) for each \utwo\ line as quadrature sum of $\Delta$(\eTeff), $\Delta$(\logg), $\Delta$(\vmicro), and $\Delta$(blend). Similarly, we determined $\Delta$(stat) for each \utwo\ line as quadrature sum of the \utwo\ line's $\Delta$(\loggf) and $\Delta$(cont). For all the stars, we list the resulting $\Delta$(sys), $\Delta$(stat), and $\Delta$(total) for each \utwo\ line in Table~\ref{tab:uncertainties}. 
\par
For the final U abundance of each sample star, we take the weighted-average of the U abundances from the three \utwo\ lines. Therefore, \epsU$ \sum_{i} (w_{i}\log \epsilon_{i})/\sum_{i} w_{i}$, where $\log \epsilon_{i}$ is the U abundance from line $i$ and $w_{i} = 1/\Delta\mathrm{(stat)}_{i}^{2}$. Here $\Delta\mathrm{(stat)}_{i}$ is the $\Delta$(stat) uncertainty as estimated above for the \utwo\ line $i$. We determined the $\Delta$(total) on the weighted-average U abundance of each star as the quadrature sum of the corresponding $\Delta$(sys) and $\Delta$(stat). For the weighted-average U abundance, the systematic uncertainty components, $\Delta$(\eTeff), $\Delta$(\logg), $\Delta$(\vmicro), and $\Delta$(blend) are determined by taking the average of these components estimated for the three \utwo\ lines. We then determined $\Delta$(sys) as the quadrature sum of the averaged $\Delta$(\eTeff), $\Delta$(\logg), $\Delta$(\vmicro), and $\Delta$(blend). For the weighted-average U abundance of each sample star, we determined $\Delta$(stat) by propagating the $\Delta$(stat) estimates of each \utwo\ for the weighted-average formula i.e., $1/\Delta\mathrm{(stat)}^{2} = \sum_{i} w_{i}$, where $w_{i} = 1/\Delta\mathrm{(stat)}_{i}^{2}$. We list the final $\Delta$(sys), $\Delta$(stat), and $\Delta$(total) for the weighted-average U abundance of each star in Table~\ref{tab:uncertainties}.
\par
We also determined systematic and statistical uncertainties for the Th and Eu abundances of all the stars. We determined $\Delta$(sys) as the quadrature sum of $\Delta$\eTeff, $\Delta$\logg, and $\Delta$\vmicro. We determined $\Delta$(stat) as the standard-error of the mean Th and Eu abundances. Therefore, $\Delta$(stat) $=\sigma/n$, where $\sigma$ is the standard deviation of the abundances determined with different lines and $n$ is the total number of lines used. We then computed $\Delta$(total) for the mean Th and Eu abundances of all the sample stars as the quadrature sum of the corresponding $\Delta$(sys) and $\Delta$(sys).

\begin{deluxetable*}{cCCCCCCCCCC}
\tablenum{6}
\tablecaption{Abundance Uncertainties \label{tab:uncertainties}}
\tablehead{
\colhead{} & \colhead{$\Delta$\eTeff\ (K)} & \colhead{$\Delta$ \logg\ (cgs)} & \colhead{$\Delta$\vmicro\ (km/s)} & \colhead{$\Delta$(blend)} & \colhead{$\Delta$(sys)}& \colhead{$\Delta$\loggf} & \colhead{$\Delta$(cont)}  & \colhead{$\Delta$(stat)} & \colhead{$\Delta$(total)}
}
\startdata 
\emhstarShort & +150 & +0.30 & +0.20 & \pm1\sigma & & & \pm0.5\% & &   \\
\hline
\eps(U)$_{3859}$ & +0.10 & +0.10 & +0.05 & \pm0.23 & \pm0.27  &\pm0.05 & \pm0.10 & \pm0.11 & \pm0.30\\
\eps(U)$_{4050}$ & +0.05 & +0.05 & +0.02 &  \pm0.30 & \pm0.31  &\pm0.03 & \pm0.10  & \pm0.10 & \pm0.33\\
\eps(U)$_{4090}$ & +0.08 & +0.09 & +0.04 &  \pm0.23 & \pm0.26  &\pm0.06 & \pm0.15  & \pm0.15 & \pm0.30\\
\eps(U) &          +0.08 & +0.08 & +0.04 & \pm0.25  & \pm0.28  &\nodata & \nodata & \pm0.07 &  \pm0.29 \\
\eps(Th) &         +0.15 & +0.08 & +0.00 & \nodata & \pm0.17 & \nodata & \nodata & \pm0.05 & \pm0.18\\
\eps(Eu) &         +0.09 & +0.07 & -0.02 & \nodata & \pm0.12 & \nodata & \nodata & \pm0.01 & \pm0.12 \\
\eps(U/Th) &       -0.07 & +0.00 & -0.00 & \pm0.25 & \pm0.26 & \nodata & \nodata & \pm0.09 & \pm0.28\\
\eps(U/Eu) &       -0.01 & +0.01 & +0.06 & \pm0.25 & \pm0.26 &  \nodata & \nodata & \pm0.07 & \pm0.27\\
\hline
\vinnistarShort & +150 & +0.30 & +0.20 & \pm1\sigma & & & \pm0.5\% & &   \\
\hline
%\eps(U)$_{3859}$ & +0.10 & +0.01 & +0.03  & \pm0.16 & \pm0.19 &\pm0.06 & \pm0.10   & \pm0.12 & \pm0.22\\
\eps(U)$_{3859}$ & +0.10 & +0.05 & +0.00  & \pm0.20 & \pm0.23 &\pm0.06 & \pm0.10   & \pm0.12 & \pm0.26\\
\eps(U)$_{4050}$ & +0.04 & +0.17 & +0.04  & \pm0.14 & \pm0.23 &\pm0.04 & \pm0.20 & \pm0.20 & \pm0.31\\
\eps(U)$_{4090}$ & +0.10 & +0.08 & +0.02  & \pm0.08 & \pm0.15 &\pm0.05 & \pm0.20   & \pm0.21 & \pm0.26\\
\eps(U) &       +0.08 & +0.10 & +0.02     & \pm0.14 & \pm0.19 & \nodata & \nodata & \pm0.09 & \pm0.21 \\
\eps(Th) &      +0.18 & +0.11 & +0.01    & \nodata & \pm0.21 &  \nodata & \nodata & \pm0.03 & \pm0.21\\
\eps(Eu) &      +0.11 & +0.07 & +0.00   & \nodata &  \pm0.13  & \nodata & \nodata & \pm0.02 & \pm0.13\\
\eps(U/Th) &    -0.10 & -0.01 & +0.01   & \pm0.14 & \pm0.17 & \nodata & \nodata & \pm0.10 & \pm0.20 \\
\eps(U/Eu) &    -0.03 & +0.03 & +0.02  & \pm0.14 & \pm0.15 & \nodata & \nodata & \pm0.09 & \pm0.17\\
\hline
\annastar & +150 & +0.30 & +0.20 & \pm1\sigma & & & \pm0.5\% & &   \\
\hline
\eps(U)$_{3859}$ & +0.10 & +0.10 & +0.03   & \pm0.08 & \pm0.17 &\pm0.04 & \pm0.06 &\pm0.07 & \pm0.18\\
\eps(U)$_{4050}$ & +0.01 & +0.25 & +0.20   & \pm0.25 & \pm0.41 &\pm0.03 & \pm0.25  &\pm0.25 & \pm0.48\\
\eps(U)$_{4090}$ & +0.10 & +0.10 & +0.05   & \pm0.11 & \pm0.19 &\pm0.05 & \pm0.21 & \pm0.22 & \pm0.28\\
\eps(U) &        +0.07 & +0.15 & +0.09     & \pm0.15 & \pm0.24 & \nodata & \nodata & \pm0.07 & \pm0.25\\
\eps(Th) &       +0.14 & +0.11 & +0.0      & \nodata & \pm0.18 & \nodata & \nodata & \pm0.06 & \pm0.19\\
\eps(Eu) &       +0.06 & +0.04 & -0.03     & \nodata & \pm0.08 & \nodata & \nodata & \pm0.02 & \pm0.08\\
\eps(U/Th) &    -0.07 & +0.04 & +0.09      & \pm0.15 & \pm0.19 & \nodata  & \nodata & \pm0.09 & \pm0.21 \\
\eps(U/Eu) &    +0.01 & +0.11 & +0.12      & \pm0.15 & \pm0.22 & \nodata  & \nodata & \pm0.07 &  \pm0.23\\
\hline
\hillstar & +150 & +0.30 & +0.20 & \pm1\sigma & & & \pm0.5\% & &   \\
\hline
\eps(U)$_{3859}$ & +0.10 & +0.10 & +0.00 & \pm0.13 & \pm0.19 & \pm0.05 & \pm0.10  & \pm0.11 & \pm0.22 \\
\eps(U)$_{4050}$ & +0.05 & +0.12 & +0.03 & \pm0.10 & \pm0.17 & \pm0.03 & \pm0.13  & \pm0.13 & \pm0.21 \\
\eps(U)$_{4090}$ & +0.10 & +0.10 & +0.00 & \pm0.07 & \pm0.16 & \pm0.06 & \pm0.18  & \pm0.19 & \pm0.25\\
\eps(U)          & +0.08 & +0.11 & +0.01 & \pm0.10 & \pm0.17 & \nodata &  \nodata & \pm0.08 & \pm0.19\\
\eps(Th) &       +0.14 & +0.06 & -0.02 & \nodata & \pm0.15 & \nodata & \nodata & \pm0.02 & \pm0.16\\
\eps(Eu) &       +0.07 & +0.08 & -0.01 &  \nodata & \pm0.11 & \nodata & \nodata & \pm0.05 & \pm0.12\\
\eps(U/Th) &     -0.06 & +0.05 & +0.03 & \pm0.10 & \pm0.13 & \nodata & \nodata & \pm0.08 & \pm0.15\\ 
\eps(U/Eu) &     +0.01 & +0.03 & +0.02 & \pm0.10 & \pm0.11 & \nodata & \nodata & \pm0.09 & \pm0.14\\
\enddata
%\tablecomments{}
\end{deluxetable*}

\section{Ages with Novel \utwo\ Lines}\label{sec:age_estimate}
The U abundance of a star can be used to determine the star's age using nucleocosmochronometry. We homogeneously determined ages for \emhstarShort, \vinnistarShort, \annastar, and \hillstar, for the first time with U abundance derived using three \utwo\ lines. We determined the ages using equations \ref{eqn:U/Th} and \ref{eqn:U/Eu} for the U/Th and U/Eu chronometers, respectively \citep{Cayrel2001_CS31082}. 
\begin{equation}\label{eqn:U/Th}
t = 21.8\rm{[log\  \epsilon(U/Th)}_{0} - log \epsilon(U/Th)_{obs}]\ Gyr
\end{equation}
\begin{equation}\label{eqn:U/Eu}
t = 14.8\rm{[log\  \epsilon(U/Eu)}_{0} - log \epsilon(U/Eu)_{obs}]\ Gyr
\end{equation}
Here \eps(U/X)$_{0}$ is the production ratio (PR) of the chronometer and \eps(U/X)$_\mathrm{obs}$ is the present-day abundance ratio of the chronometer as determined from this work. We took PRs from \cite{Schatz2002}, who used waiting-point calculations to estimate site-independent PRs of \rproc\ elements. We list the final ages from the two chronometers and the associated uncertainties in Table~\ref{tab:Abundances}. We find the U/Th and U/Eu age agreeing for all the stars within uncertainties. There is a relatively large discrepancy between the U/Th and U/Eu ages of \hillstar, which can be attributed to its actinide-boost nature.
\par
We estimated the uncertainty on the ages by propagating the uncertainties of the PRs and the present-day observed abundance ratios. As a result, our age uncertainties consist of statistical, systematic, and PR components. We obtained the uncertainty on the PRs from \cite{Schatz2002}, which are $0.10$ dex for the U/Th chronometer and $0.11$ dex for the U/Eu chronometer. We determined $\Delta$(sys) for the present-day abundance ratios as the quadrature sum of the $\Delta$\eTeff, $\Delta$\logg, $\Delta$\vmicro, $\Delta$(blend) uncertainties of the ratios. We determined $\Delta$(stat) for the present-day abundance-ratios as quadrature sum of $\Delta$(stat) determined for the elements in the ratio. We list the resulting $\Delta$(sys) and $\Delta$(stat) for the present-day U/Th and U/Eu abundance ratios in Table~\ref{tab:uncertainties}. We list the individual systematic, statistical and production-ratio components of the age uncertainties in Table~\ref{tab:Abundances}. We determined the total uncertainty on the age as the quadrature sum of the systematic, statistical, and production-ratio uncertainties. We further discuss the ages and the respective uncertainties from the various components in Section~\ref{subsec:discuss_ages}.

\begin{figure*}
    \centering
    \includegraphics[width = \textwidth]{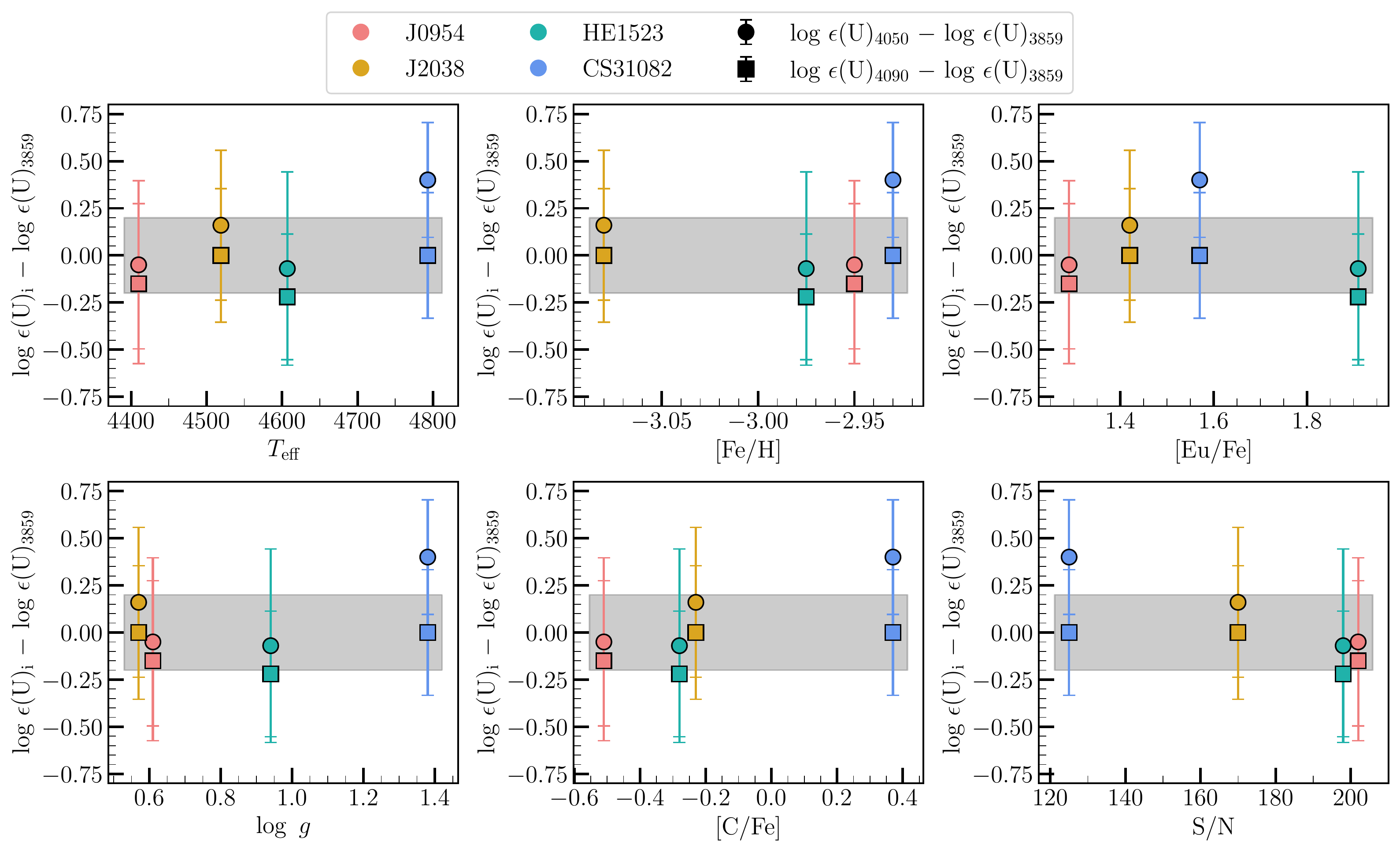}
    \caption{For all the sample stars, residuals between the $\lambda4050$\,\AA\ and $\lambda3859$\,\AA\ \utwo\ line abundances are shown with round data points and the residuals between the $\lambda4090$\,\AA\ and $\lambda3859$\,\AA\ \utwo\ line abundances are shown with square data points. The residuals are plotted against the \eTeff, \logg, [Fe/H], [C/Fe], and S/N at 4050\,\AA\ of the respective stars in different panels. A grey-shaded region for residuals within $\pm\ 0.2$ dex is also shown.}
    \label{fig:delU3859}
\end{figure*}

\section{Discussion}\label{sec:discussion}
To test the reliability of the two new \utwo\ lines at $\lambda4050$ and $\lambda4090$\,\AA, we performed the first homogeneous U abundance analysis of four highly RPE stars ([Eu/Fe]$ > +0.7$) with these new lines, in addition to the canonical $\lambda3859$\,\AA\ \utwo\ line. The stars chosen for the analysis are four of the five stars with U abundance previously determined in the literature. While U abundance was determined for \cs\ \citep{Hill2017_cs29497}, our analysis of the star's UVES/VLT spectra indicated an almost-non existent signature of U at all three \utwo\ lines. Therefore, we left \cs\ out of further analysis. We now discuss and establish the reliability of the two new \utwo\ lines in the other four stars.

\subsection{Reliability of the New \utwo\ lines: Line Abundances and Uncertainties}\label{subsec:discuss_relabel1}
Table~\ref{tab:Abundances} lists the final U abundance determined at each \utwo\ line, along with its estimated uncertainty, for all the sample stars. We note that the U abundances from the three \utwo\ lines agree well, within uncertainties, for all the sample stars. We also find that the $\lambda3859$ U abundance determined in this work is consistent with previous literature estimate of the same within uncertainties for all the stars; $\lambda3859$ U abundance from literature is also listed in Table~\ref{tab:Abundances}. 
\par
Moreover, we find that in the case for all the sample stars, the uncertainties on the U abundances of all three \utwo\ lines are of the same order. This indicates that the new \utwo\ lines provide similar precision for U abundance determination as the canonical $\lambda3859$\,\AA\ \utwo\ line. For all three \utwo\ lines, we have homogeneously taken into account various sources of systematic and statistical uncertainties (see section~\ref{subsec:uncertainty}). Generally, we find the systematic uncertainties, specifically, from the blending elements to be the dominant source of the total uncertainty on the \utwo\ line abundances. We also find that U abundance determination with all three \utwo\ lines is sensitive to the continuum placement of the synthetic model. The sensitivity of the individual \utwo\ lines to the various sources of uncertainties and the similar precision offered by the \utwo\ lines impresses upon the advantage of using three \utwo\ lines for U abundance determination, instead of just one.
\par
We note, however, an exceptionally high uncertainty estimated for the $\lambda4050$ U abundance of \annastar, on the order of $\sim0.5$ dex. This high estimate is driven by the \ion{La}{2} blend since the star has a relatively strong La feature relative to the other stars (see Figure~\ref{fig:4050EditedZoomIn}). This does highlight the fact that blends can have a significant impact on U abundance determination. However, we find that the spectral synthesis fit of the region is very good for the derived La abundance of this star. 
\par
As noted in section~\ref{subsec:U}, even though the spectral synthesis fits to the \utwo\ lines are good --- the goodness of the fits is further corroborated by the agreement between U abundances from the different \utwo\ lines --- there is indication of unidentified features in the $\lambda4090$\,\AA\ and possibly $\lambda4050$\,\AA\ spectral regions. Given the immense potential of these \utwo\ lines on advancing nucleocosmochronometry and \rproc\ studies, we recommend a detailed investigation into the atomic data, especially laboratory measurements of the transition lines in these spectral regions. 

\begin{figure}
    \centering
    \includegraphics[width = 0.47\textwidth]{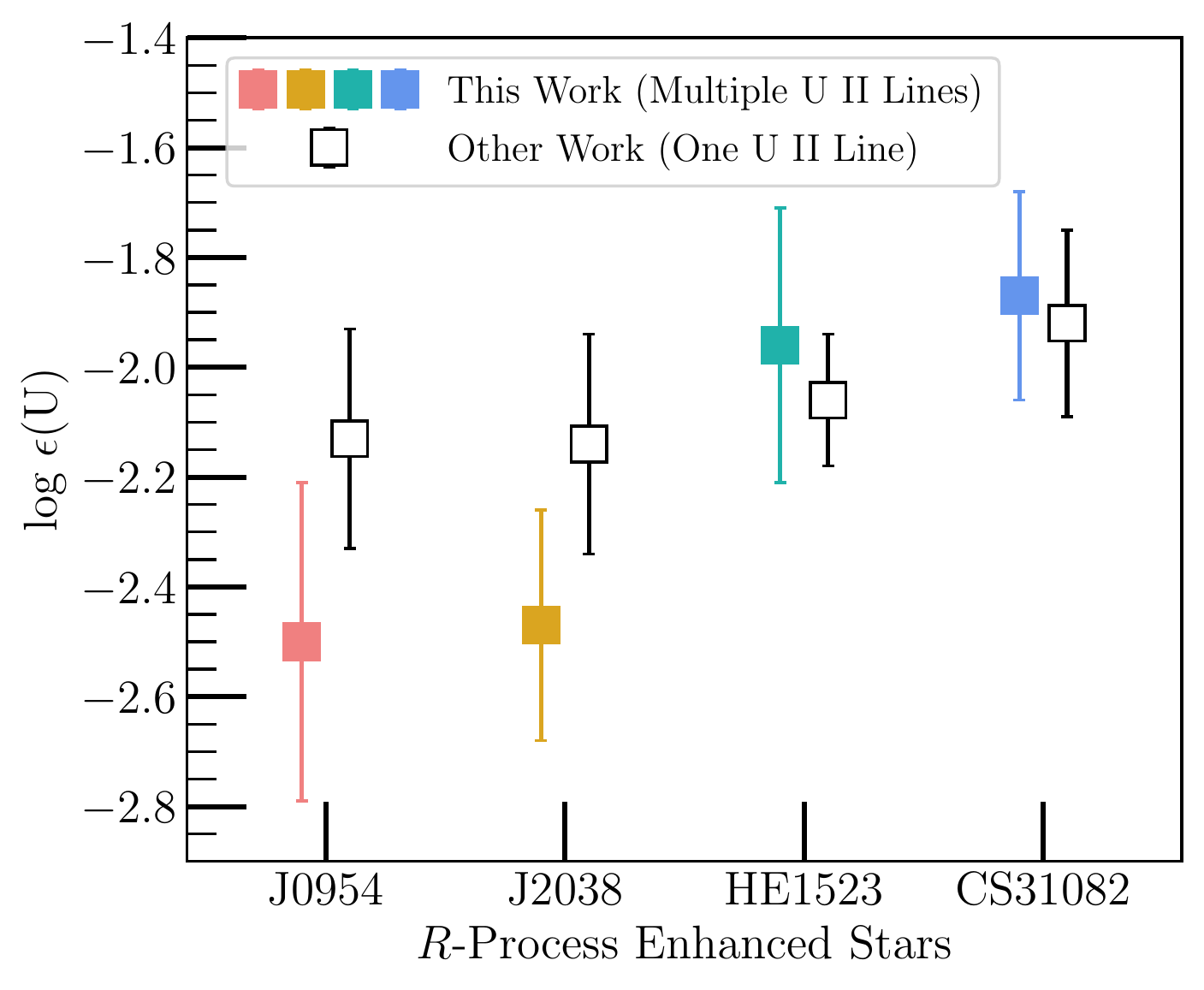}
    \caption{Weighted-average U abundances derived in this work with \utwo\ lines at $\lambda3859$, $\lambda4050$, and $\lambda4090$\,\AA\ for \emhstarShort, \vinnistarShort, \annastar, and \hillstar. U abundances determined in previous literature studies using the $\lambda3859$\,\AA\ line are also shown. Literature U abundance was taken from \cite{holmbeckJ0954_2018} for \emhstarShort, \cite{PlaccoJ2038_2017} for \vinnistarShort, \cite{FrebelHE1523_2007} for \annastar, and \cite{HillCS31082_2002} for \hillstar.}
    \label{fig:MeanStars}
\end{figure}
\subsection{Reliability of the New \utwo\ lines: Residuals between Line Abundances}\label{subsec:discuss_reliable2}
We further demonstrate the reliability of the new \utwo\ lines with Figure~\ref{fig:delU3859}, which shows (i) the residuals between the $\lambda4050$ and $\lambda3859$ U abundances in circle data points and (ii) the residuals between the $\lambda4090$ and $\lambda3859$ U abundances in square data points, for all the sample stars. The residuals are plotted against the respective star's \eTeff, \logg, [Fe/H], [C/Fe], [Eu/Fe], and spectrum S/N in different panels. The uncertainties on the residual data points are calculated by propagating the total uncertainties of the individual line abundances. 
\par
First, we note that the residuals of the U abundances are all within $\pm0.2$ dex (shown by the shaded grey region) for all the stars. For \hillstar, while we find that the residual between the $\lambda4090$ and $\lambda3859$ U abundances is $0.0$ dex, the residual between the $\lambda4050$ and $\lambda3859$ U abundances is $+0.40$ dex. This indicates that there might be an unidentified transition line in the $\lambda4050$\,\AA\ \utwo\ line region that is more prominent for \hillstar\ than for the other stars. Alternatively, the abundance of the \ion{La}{2} HFS structure in this region may not be well represented by the mean La abundance determined for the star. While this relatively large residual may signify the need to better constrain the atomic data of this spectral region and/or the abundance of the \ion{La}{2} HFS structure, the uncertainty on the residual overlaps with the $\pm0.2$ dex shaded-region, subduing any serious concern. 
\par
Second, we note that the residuals of the U abundances show no discernible trend with respect to \eTeff, \logg, [Fe/H], [C/Fe], [Eu/Fe], and spectrum S/N. This indicates that our current spectral synthesis models are of high fidelity and no identifiable systematic biases are being percolated to the U abundance determinations. Together, the $\pm0.2$ dex range of the residuals between the \utwo\ line abundances and the absence of any significant trend in the residuals with respect to key atmospheric and chemical properties as well as data-quality of the sample stars establishes the reliability of the two new \utwo\ new lines at $\lambda4050$\,\AA\ and $\lambda4090$\,\AA. 

\subsection{Mean U Abundance with Multiple \utwo\ lines}\label{subsec:discuss_mean}
We provide revised U abundances for the RPE stars, \emhstarShort, \vinnistarShort, \annastar, and \hillstar\ via a homogeneous analysis of the three \utwo\ lines at $\lambda3859$\,\AA, $\lambda4050$\,\AA, and $\lambda4090$\,\AA. The resulting mean U abundance is computed as a weighted-average of the U abundances from the three \utwo\ lines, with the weights assigned based on the statistical uncertainty on the U abundance of each line (see section~\ref{subsec:U} for more details). We find that the total uncertainty of the weighted-average U abundances is on the order of $\sim0.2$ dex, for all sample stars. On the other hand, the total uncertainty of individual \utwo\ line abundances is higher and typically in the range of $\sim0.2-0.3$ dex. This reiterates the advantage of using multiple \utwo\ lines, which can potentially lend more precise U abundance than a single \utwo\ line. 
\par
We compare the final weighted-average U abundances estimated in this work to previous literature estimates of $\lambda3859$ U abundances in Figure~\ref{fig:MeanStars}, for all the stars. We find that our U abundances agree well with the literature values within uncertainties, for all the stars. In fact, the final U abundances agree with previous literature estimates within $0.05$ dex for \annastar\ and \hillstar, which we consider excellent agreement. This further establishes the reliability of the new \utwo\ lines. For \emhstarShort, we find that our U abundance is $\sim0.4$ dex lower than that determined by \cite{holmbeckJ0954_2018}. We investigated the source of this discrepancy and suspect the cause to be differences in the adopted atomic data of the $\lambda$3859.91\,\AA\ \ion{Fe}{1} line. We also find that our U abundance for \vinnistarShort\ is $\sim0.3$ dex lower than that determined by \cite{PlaccoJ2038_2017}. This discrepancy is attributed to the difference in the adopted stellar parameters, especially \logg\ (see section \ref{sec:stellarparams}). Nevertheless, our U/Th and U/Eu nucleocosmochronometric age estimates compare well with those reported in \cite{PlaccoJ2038_2017} (see section \ref{sec:age_estimate}). 
\par
We also discuss the uncertainty estimates on the final U abundances as determined in this work compared to previous literature estimates. In the case of \emhstarShort\ and \annastar, the total uncertainty estimated for the final weighted-average U abundance in this work is larger than the uncertainty quoted by the respective previous literature studies for their final $\lambda3859$ U abundance (see Figure~\ref{fig:MeanStars} and/or Table~\ref{tab:Abundances}). For \emhstarShort, \cite{holmbeckJ0954_2018} quoted a fiducial uncertainty of $\pm0.20$ dex on their final U abundance, while we obtained an uncertainty of $\pm0.29$ dex. For \annastar, \cite{FrebelHE1523_2007} assigned an uncertainty of $\pm0.11$ dex, solely arising from the effect of changing the Fe abundance by $\pm0.10$ dex (although they considered additional sources of uncertainties in the age determinations). Having accounted for additional sources of uncertainties, we obtained a larger uncertainty of $\pm0.25$ dex for the U abundance of \annastar. While \cite{PlaccoJ2038_2017} also set a fiducial uncertainty of $\pm0.20$ dex for \vinnistarShort, we find our detailed analysis renders a similar uncertainty of $\pm0.21$ dex. Similarly for \hillstar, we find that our U abundance uncertainty of $\pm0.19$ dex agrees with that of \cite{HillCS31082_2002}, who also accounted for stellar parameters, oscillator strength, and observational fitting uncertainties and obtained an uncertainty of $\pm0.19$ dex.

\begin{figure}
    \centering
    \includegraphics[width = 0.47\textwidth]{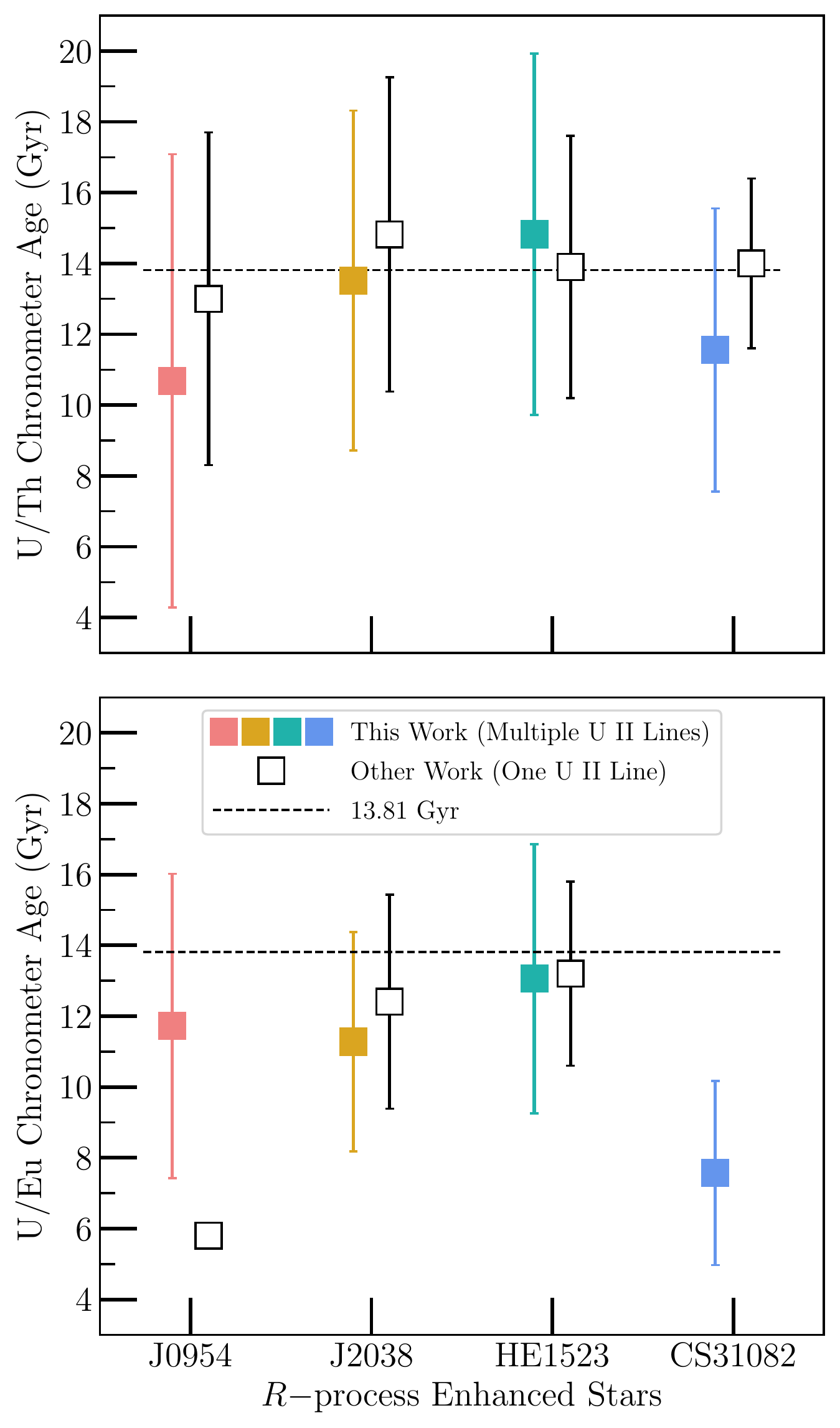}
    \caption{Nucleocosmochronometric ages from this work (colored data points) and previous literature work (white data points) using the U/Th (top panel) and U/Eu (bottom panel) chronometers. The age of the universe is shown in dashed-black line \citep{Planck2016_age13.8}. For the ages of this work, the weighted-average U, Th, and Eu abundances from multiple transition lines were used. Literature ages were taken from \cite{holmbeckJ0954_2018} for \emhstarShort, \cite{PlaccoJ2038_2017} for \vinnistar, \cite{FrebelHE1523_2007} for \annastar, and \cite{HillCS31082_2002} for \hillstar.}
    \label{fig:AgesThAvg}
\end{figure}
\subsection{Ages with Mean U Abundances}\label{subsec:discuss_ages}
For every sample star, we determined two stellar ages, one from the U/Th chronometer and another using the U/Eu chronometer. We list the resulting ages in Table~\ref{tab:Abundances}. There is a good agreement between the U/Th and U/Eu ages of \emhstarShort, \vinnistarShort, and \annastar. For \hillstar, U/Eu age is $\sim4.0$ Gyr lower than the U/Th age due to its actinide-boost nature \citep{Cayrel2001_CS31082, HillCS31082_2002, Schatz2002}. Even then, the U/Th and U/Eu ages of \hillstar\ agree within uncertainties. 
\par
For stellar age uncertainties, we took into account systematic, statistical, and PR uncertainties as described in section~\ref{sec:age_estimate}. These individual age uncertainty components are also listed in Table~\ref{tab:Abundances}, along with the total age uncertainties. For the U/Th and U/Eu stellar ages of all the sample stars, the systematic uncertainties are the largest (and also the most dominant, in many cases), followed by the PR uncertainties and then the statistical uncertainties. Specifically, the systematic uncertainties of the stellar ages are driven by the large U abundance uncertainties from the blending elements.
\par
Additionally, the uncertainties on the U/Th stellar ages are larger than the U/Eu counterpart because of the longer half-life of Th, relative to U. We note that we have not taken into account the systematic uncertainties associated with using just one set of theoretical PRs \citep[from][]{Schatz2002}. Other \rproc\ models have predicted slightly varying zero-age abundance ratios for U/Th and U/Eu \citep[e.g.,][]{Goriely2001_multieventrprocmodel, Farouqu2010_highentropywind}. However, since the aim  of this study was to investigate stellar ages estimated with revised U abundances from multiple \utwo\ lines, we find using one set of PRs sufficient for the analysis.
%xx standard-deviation on these xx. These depict that ultimately U/Eu maybe be only as accurate as U/Th, if not less. 
\par
We compare all the stellar ages determined in this work to previous literature results in Figure~\ref{fig:AgesThAvg}, which shows the U/Th and U/Eu ages in the top and bottom panels, respectively. The dashed black line in Figure~\ref{fig:AgesThAvg} indicates the age of the universe, as determined by the \textit{Planck} mission \citep{Planck2016_age13.8}. For the literature ages of \emhstarShort\ \citep{holmbeckJ0954_2018}, \vinnistarShort\ \citep{PlaccoJ2038_2017}, and \annastar\ \citep{FrebelHE1523_2007}, we display the ages determined by the respective studies using, specifically, the zero-age abundance ratios of \cite{Schatz2002} to facilitate a consistent comparison to our age-estimates. For \hillstar, \cite{HillCS31082_2002} determined U/Th age using zero-age abundance ratio from \cite{Goriely2001_multieventrprocmodel}, which we display. Also, \cite{HillCS31082_2002} did not determine the U/Eu stellar age of the \hillstar. 
\par
We find that our stellar ages mostly agree well with previous literature results within uncertainties. The exception to this case is the U/Eu age of \emhstarShort, which is much higher than previously determined in the literature. We attribute this discrepancy to a lower U abundance determined in this work relative to that determined in \cite{holmbeckJ0954_2018} (see section~\ref{subsec:discuss_mean} more details). We also determined lower Th abundance relative to \cite{holmbeckJ0954_2018} so that in the U/Th ratio, the offset in the abundances is cancelled.

\section{Conclusions}\label{sec:conclusions}
Uranium abundances of metal-poor RPE stars enable stellar age determination independent from stellar evolution models, using nucleocosmochronometry. U abundances of a large sample of RPE stars may also enable important constraints on the astrophysical conditions and nuclear physics of \rproc\ enrichment events, by probing the production of the actinides. However, U abundance determination has been limited to using a single \utwo\ line at $\lambda3859$\,\AA. In this study, we have performed the first homogeneous U abundance analysis of four highly RPE stars with two new \utwo\ lines at $\lambda4050$\,\AA\ and $\lambda4090$\,\AA, along with the canonical $\lambda3859$\,\AA\ \utwo\ line.
\par
We test the utility of the $\lambda4050$\,\AA\ and $\lambda4090$\,\AA\ \utwo\ lines and find them generally reliable for U abundance determination. In Figures \ref{fig:4050EditedZoomIn} and \ref{fig:4090ZoomIn}, we show the spectral synthesis fits to the new \utwo\ lines. The resulting $\lambda3859$, $\lambda4050$, and $\lambda4090$ U abundances agree with each other within $\pm0.2$ dex for most stars, and within uncertainties for all the stars, as seen in Figure~\ref{fig:delU3859}. In fact, we find it particularly advantageous to use the $\lambda4050$\,\AA\ and $\lambda4090$\,\AA\ \utwo\ lines, since they are not blended with strong lines or C features. In that regard, they may find a special utility for determining U abundances of C-enhanced metal-poor stars. As seen in Figure~\ref{fig:MeanStars}, the final weighted-average U abundances of all the sample stars agree with previous literature estimates. This substantiates our U abundance analysis framework and the use of multiple \utwo\ lines. 
\par
We performed a detailed uncertainty analysis of the U abundances by taking into account systematic uncertainties from stellar parameters and blends, as well as statistical uncertainties from continuum placement and \loggf\ measurements. We find that all three \utwo\ lines provide similar precision of $\sim 0.2-0.3$ dex. On the other hand, for the weighted-average U abundance, the uncertainties are on the order of $\sim 0.2$ dex. This underscores the advantage of using multiple \utwo\ lines. Moreover, any unconstrained systematic biases associated with a particular \utwo\ line are mitigated in the average abundance from multiple \utwo\ lines. 
\par
We also obtained homogeneous ages for the stars with the U/Th and U/Eu chronometers and using the U, Th, and Eu abundances as derived in this work from multiple transition lines of each element. As seen in Figure~\ref{fig:AgesThAvg}, we find that the newly obtained ages are reasonable and in agreement with previous literature estimates within uncertainties. For the uncertainties on the ages, we estimated the systematic, statistical, and PR uncertainty components for all the stars. The resulting total uncertainties on the ages of the sample stars are on the order of $\sim 4-6$ Gyr for the U/Th ages and $\sim 3-4$ Gyr for the U/Eu ages. 
\par
To improve the uncertainties on the U abundances and subsequently stellar ages, it will be necessary to address systematic uncertainties from the blends and stellar parameters. Additionally, a substantial component of U abundance uncertainties is contributed to fitting uncertainties like continuum placement. As a result, studies of these new \utwo-line spectral regions are recommended to better constrain the atomic parameters of the blends and to identify unknown neighboring transitions, which will improve the confidence in the continuum placement. 
\par
Another source of uncertainty on the stellar ages is the poorly known nuclear physics that enters into the predicted U/Th and U/Eu PRs. Upcoming studies at facilities such as the N=126 Factory at Argonne National Laboratory \citep{Savard+2020} and the Facility for Rare Isotope Beams \citep{Castelvecchi2022} will reach many heavy, neutron-rich species whose properties are crucial for understanding actinide production. These anticipated advances in atomic and nuclear physics will contribute to the overarching goal of improving the precision of nucleocosmochronometry.
\par
Through the rest of the decade, we are expecting an influx of new spectroscopic data, specifically for RPE stars, from surveys such as that by the $R$-Process Alliance \citep{rpa1, rpa2, rpa3, rpa4}, 4MOST \citep{4MOST2019}, and WEAVE \citep{WEAVE2012}. Reliable U abundance determination of RPE stars will be critical in obtaining precise and robust nucleocosmochronometric ages of some of the oldest stars. Precise nucleocosmochronometric ages, combined with chemo-dynamical information, can aid our understanding of the chemical enrichment and evolution in the early Universe, especially of the \rproc\ elements, as well as the assembly history of our galaxy. Additionally, reliable U abundances for a large sample of RPE stars can shed light on the extent of actinide-variation in RPE stars and its origin. To that end, the results of this work open up a new avenue to reliably determine U abundances and nucleocosmochronometric ages for a large sample of RPE stars with multiple \utwo\ lines. 

%The last numbered section should briefly summarise what has been done, and describe
%the final conclusions which the authors draw from their work.

\begin{acknowledgments}
SPS acknowledges Jamie Tayar for useful conversations on stellar ages and other comments. RE acknowledges support from NSF grant AST-2206263.
APJ was supported by NASA through Hubble Fellowship grant HST-HF2-51393.001 awarded by the Space Telescope Science Institute, which is operated by the Association of Universities for Research in Astronomy, Inc., for NASA, under contract NAS5-26555. APJ also acknowledges support from a Carnegie Fellowship and the Thacher Research Award in Astronomy.
T.T.H acknowledges support from the Swedish Research Council (VR 2021-05556).
This project initiated during MC's 2018 sabbatical stay at Carnegie Observatories, and he is very grateful to the faculty and staff there for their hospitality and generous support. Additional support for MC is provided by the Ministry for the Economy, Development, and Tourism's Millennium Science Initiative through grant ICN12\textunderscore 12009, awarded to the Millennium Institute of Astrophysics (MAS), and by Proyecto Basal CATA ACE210002 and FB210003.
I.U.R.\ acknowledges support from the U.S.\ National Science Foundation (NSF) (grants PHY~14-30152---Physics Frontier Center/JINA-CEE, AST~1815403/1815767, and AST~2205847), and the NASA Astrophysics Data Analysis Program, grant 80NSSC21K0627.
EMH acknowledges support for this work provided by NASA through the NASA Hubble Fellowship grant HST-HF2-51481.001 awarded by the Space Telescope Science Institute, which is operated by the Association of Universities for Research in Astronomy, Inc., for NASA, under contract NAS5-26555.
T.C.B. acknowledges partial support for this work from grant PHY 14-30152; Physics Frontier Center/JINA Center
for the Evolution of the Elements (JINA-CEE), awarded by the US National Science Foundation. This work has made use of data from the European Space Agency (ESA) mission {\it Gaia} (\url{https://www.cosmos.esa.int/gaia}), processed by the {\it Gaia} Data Processing and Analysis Consortium (DPAC, \url{https://www.cosmos.esa.int/web/gaia/dpac/consortium}). Funding for the DPAC has been provided by national institutions, in particular the institutions participating in the {\it Gaia} Multilateral Agreement. The authors wish to recognize and acknowledge the very significant cultural role and reverence that the summit of Maunakea has always had within the indigenous Hawaiian community.  We are most fortunate to have the opportunity to conduct observations from this mountain. 
\end{acknowledgments}


\vspace{5mm}
\facilities{Keck/HIRES, Magellan/MIKE, VLT/UVES}


\software{\texttt{astropy} \citep{Astropy2013},  
          %Cloudy \citep{2013RMxAA..49..137F}, 
          %Source Extractor \citep{1996A&AS..117..393B}
          \texttt{MAKEE} (\url{https://sites.astro.caltech.edu/~tb/makee/}),
          \texttt{CarPy} \citep{Kelson2000_CarPy1stref, Kelson2003_CarPy2ndref},
          \texttt{ESOReflex} \citep{Freudling2013_ESOReflex},
          \texttt{MOOG} (\url{https://github.com/alexji/moog17scat} and \cite{Sneden1973_MOOG}),
          \texttt{SMHr} (\url{https://github.com/eholmbeck/smhr-rpa/tree/refactor-scatterplot} and \url{https://github.com/andycasey/smhr/tree/refactor-scatterplot})
          }



\appendix
\section{Hyperfine Splitting of the \ion{La}{2} $\lambda$4050 Line}\label{Appendix_la4050hfs}

The U~\textsc{ii} line at 4050.04~\AA\ is blended
with a stronger La~\textsc{ii} line at 4050.073~\AA.~
There is one dominant naturally occurring isotope of La,
$^{139}$La, which has nuclear spin $I = 7/2$.
This non-zero nuclear spin creates HFS
structure, which desaturates the line and is
thus important to account for in stellar abundance work.
We adopt the HFS $A$ and $B$ constants for the
upper and lower levels of this transition from the
measurements of
\citet{furmann08a,furmann08b}.
We compute the complete line component pattern for this
line following the procedure described in Appendix~A1 of
\citet{ivans06}.
We calculate the center-of-gravity wavenumber of this line
from the energy levels given in the
National Institute of Standards and Technology (NIST)
Atomic Spectra Database (ASD; \citealt{kramida21}).
We convert to the center-of-gravity air wavelength using the 
standard index of air \citep{peck72}.
Table~\ref{lahfstab} lists the line component positions
relative to these values.
The strengths are normalized to sum to 1.0.

\begin{deluxetable}{ccccccc}
\tablecaption{Hyperfine Structure
Line Component Pattern for the La~\textsc{ii} $\lambda$4050 Line
\label{lahfstab}}
%\tablewidth{0pt}
\tablenum{A1}
\tabletypesize{\small}
\tablehead{
\colhead{Wavenumber} &
\colhead{$\lambda_{\rm air}$} &
\colhead{$F_{\rm upper}$} &
\colhead{$F_{\rm lower}$} &
\colhead{Component Position} &
\colhead{Component Position} &
\colhead{Strength} \\
\colhead{(cm$^{-1}$)} &
\colhead{(\AA)} &
\colhead{} &
\colhead{} &
\colhead{(cm$^{-1}$)} &
\colhead{(\AA)} &
\colhead{}
}
\startdata
24683.94 & 4050.073 & 5.5 & 6.5 &  $+$0.188754 & $-$0.030979 & 0.25000 \\
24683.94 & 4050.073 & 5.5 & 5.5 &  $+$0.097552 & $-$0.016010 & 0.04545 \\
24683.94 & 4050.073 & 5.5 & 4.5 &  $+$0.018860 & $-$0.003095 & 0.00455 \\
24683.94 & 4050.073 & 4.5 & 5.5 &  $+$0.066181 & $-$0.010862 & 0.16883 \\
24683.94 & 4050.073 & 4.5 & 4.5 & $-$0.012510 &  $+$0.002053 & 0.06926 \\
24683.94 & 4050.073 & 4.5 & 3.5 & $-$0.077930 &  $+$0.012790 & 0.01190 \\
24683.94 & 4050.073 & 3.5 & 4.5 & $-$0.037194 &  $+$0.006104 & 0.10476 \\
24683.94 & 4050.073 & 3.5 & 3.5 & $-$0.102614 &  $+$0.016842 & 0.07483 \\
24683.94 & 4050.073 & 3.5 & 2.5 & $-$0.154142 &  $+$0.025298 & 0.02041 \\
24683.94 & 4050.073 & 2.5 & 3.5 & $-$0.121201 &  $+$0.019892 & 0.05612 \\
24683.94 & 4050.073 & 2.5 & 2.5 & $-$0.172729 &  $+$0.028349 & 0.06531 \\
24683.94 & 4050.073 & 2.5 & 1.5 & $-$0.209879 &  $+$0.034446 & 0.02857 \\
24683.94 & 4050.073 & 1.5 & 2.5 & $-$0.185678 &  $+$0.030474 & 0.02143 \\
24683.94 & 4050.073 & 1.5 & 1.5 & $-$0.222828 &  $+$0.036572 & 0.04286 \\
24683.94 & 4050.073 & 1.5 & 0.5 & $-$0.245257 &  $+$0.040253 & 0.03571 \\
\enddata
%\tabletypesize{\scriptsize}
\tablecomments{%
Energy levels from the NIST ASD and the index of air \citep{peck72}
are used to compute the
center-of-gravity wavenumbers and air wavelengths, $\lambda_{\rm air}$.
Line component positions are given relative to those values.
Component strengths are normalized to sum to 1.0.
}
\end{deluxetable}

\bibliography{main}{}
\bibliographystyle{aasjournal}



\end{document}